\documentclass[acmsmall,natbib=false]{acmart}

\usepackage{float}
\usepackage{comment}
\usepackage{subfig}
\usepackage{caption}
\usepackage{graphicx,xcolor} 
\usepackage[framemethod=tikz]{mdframed}

\usepackage{algorithm, algpseudocode}
\usepackage{amsmath}

\AtBeginDocument{%
  }

\setcopyright{acmcopyright}
\copyrightyear{2022}
\acmYear{2022}
\acmDOI{XXXXXXX.XXXXXXX}

\acmJournal{JACM}

\RequirePackage[
  datamodel=acmdatamodel,
  style=acmauthoryear,
  ]{biblatex}

\begin{document}

\title{EVHA: Explainable Vision System for Hardware Testing and Assurance - An Overview}

\author{Md Mahfuz Al Hasan}
\email{mdmahfuzalhasan@ufl.edu}
\author{Mohammad Tahsin Mostafiz}
\email{m.tahsinmostafiz@ufl.edu}
\author{Thomas An Le}
\email{thomas.le@ufl.edu}
\author{Jake Julia}
\email{jake.julia@ufl.edu}
\author{Nidish Vashistha}
\email{nidish@ufl.edu}
\author{Shayan Taheri}
\email{shayan.taheri@ufl.edu}
\author{Navid Asadizanjani}
\email{nasadi@ece.ufl.edu}
\affiliation{%
  \\
  \institution{Florida Institute for Cybersecurity Research, University of Florida}
  \streetaddress{300 SW 13TH ST, P.O. Box 113150}
  \city{Gainesville}
  \state{Florida}
  \country{USA}
  \postcode{32601}
}

\renewcommand{\shortauthors}{Al Hasan et al.}

\begin{abstract}
  Due to the ever-growing demands for electronic chips in different sectors the semiconductor companies have been mandated to offshore their manufacturing processes. This unwanted matter has made security and trustworthiness of their fabricated chips concerning and caused creation of hardware attacks. In this condition, different entities in the semiconductor supply chain can act maliciously and execute an attack on the design computing layers, from devices to systems. Our attack is a hardware Trojan that is inserted during mask generation/fabrication in an untrusted foundry. The Trojan leaves a footprint in the fabricated through addition, deletion, or change of design cells. In order to tackle this problem, we propose Explainable Vision System for Hardware Testing and Assurance (EVHA) in this work that can detect the smallest possible change to a design in a low-cost, accurate, and fast manner. The inputs to this system are Scanning Electron Microscopy (SEM) images acquired from the Integrated Circuits (ICs) under examination. The system output is determination of IC status in terms of having any defect and/or hardware Trojan through addition, deletion, or change in the design cells at the cell-level. This article provides an overview on the design, development, implementation, and analysis of our defense system.
\end{abstract}



\maketitle

\section{Introduction}\label{sec:introduction}

The globalization of the semiconductor design, fabrication, assembly, and test has increased concerns over the trustworthiness of Integrated Circuits (ICs) and systems and they have become vulnerable to malicious activities and alterations. There are a number of entities in the semiconductor supply chain, including third party intellectual property vendor, system integrator, foundry, assembly and packaging, test facility, and end-user, that can act as malicious actor. These entities can be engaged in the provision of attack mechanisms at any computing layer, from devices to systems. Hence, establishing assurance and trust in ICs has become of utmost importance \cite{rostami2014primer, rostami2013hardware, nozawa2021generating, zhao2019density, yasaei2021gnn4tj, shi2019golden, chen2017single, hasegawa2017trojan, yu2017exploiting}. Among these entities, however, “untrusted foundries” has received significant attention over the past decade or so, as they have complete access to the design file and can maliciously manipulate the circuit.

The significant attentions toward this entity in the IC supply chain are due to: (a) existence of few advanced foundries around the globe according to which majority of them are considered as untrusted; and (b) provision of complete access into design details to foundries that makes them capable of maliciously changing the circuit in a manner that would be very difficult to detect. In our threat model, the adversary injects a hardware Trojan (i.e., small malicious modifications on IC) into the design circuit (i.e., gate/cell-level Trojan) during mask generation/fabrication (in foundry) through addition, deletion, or change of design cells. The inserted Trojan leaves a footprint in the fabricated chip. Meanwhile, the design file known as Graphic Design System for Information Interchange (GDSII) is considered to be trusted in this scenario.

Although there is a flurry of investments and activities to address the untrusted foundry problem and bring semiconductor manufacturing on-shore, the problem still remains. Example initiatives include the Creating Helpful Incentives to Produce Semiconductors (CHIPS) for America Act legislation from U.S. with \$52 billion investment \cite{IntelAmerica_2021}, the China Integrated Circuit Investment Industry Fund (CICIIF) from China with \$150 billion investment \cite{SutterChina_2021}, K-Semiconductor Belt from South Korea with \$450 billion investment \cite{AlamChip_2021}, along with TSMC’s new initiatives in Arizona \cite{TSMC_2021} and Intel’s new fabrication initiatives \cite{IntelAmerica_2021}. Although increasing the number of fabrication facilities (Integrated Device Manufacturers (IDM) and pure play foundries) around the globe will help address the supply chain issues and silicon shortage problem exacerbated by COVID-19 \cite{PitkinChip_2021}, the trust concerns will remain as chip fabrication is going to remain a global effort with electronic systems built using ICs of different companies fabricated by different foundries. Making it worse, as the number of transistors per chip continues to grow, attacks have shown that adversary can design ever more intelligent hardware Trojans with fewer and fewer gates, making them extremely difficult to detect using traditional test and side channel techniques \cite{li2016survey}.

Hardware Trojans can seriously degrade the functionality, performance, and reliability of electronic systems. They can cause loss of profit for electronic devices to life-threatening payloads for all types of applications including transportation, healthcare, and military systems. The detection approaches for hardware Trojans are categorized as \cite{bao2014application}: (a) Test-Time, in which post-silicon tests are applied. It is generally divided into Functional/Structural Test and Side-Channel Fingerprinting. These methods have serious shortcomings with respect to their detection effectiveness as Trojans are hard to model and could be designed to stay dormant during chip operation. (b) Run-Time, in which a circuitry is added for monitoring the behavior/state of a chip after it has been deployed into the field. The major disadvantages include high area overhead and that the run-time detection circuitry is Trojan-free. Both of the mentioned categories require a golden model (intended functionality and behavior) to compare with. (c) Reverse-Engineering, in which the reverse engineering process is applied to the ICs in order to detect the Trojans. The recovered design is compared to a golden netlist to determine its trustworthiness. The major shortcomings include very high cost, fully destructive operation, requires large samples of ICs, and that it is a very time-consuming process that could take months for a successful effort.

Unfortunately, majority of these techniques are only applicable to digital circuits but none could be used for analog and mixed signal devices. Currently, there is a lack of efficient techniques to detect malicious implants (digital or analog, with or without payload, noisy or dormant, small or large, etc.) in different types of integrated circuits (analog/mixed-signal, digital, memory, etc.). Among the many techniques investigated thus far, it is believed that non-invasive or semi-invasive physical inspection is the most promising for rapid detection of malicious change to an IC, even as small as a single transistor, at a lower cost, by an untrusted foundry when GDSII is available. There are limitations for developing efficient physical inspection-based detection such as (1) Rapid increase in the number of transistors in the chip; (2) smaller Trojan sizes inserted in the chip; (3) increasingly stealthy Trojans; (4) unique and unexplored nature of IC Scanning Electron Microscopy (SEM) images due to strong repetitive features, process variation effects, and the lack of sufficient images in different technology nodes and vendors.

Recent advancements in deep learning from Artificial Intelligence (AI) and computational resources for object detection and segmentation/recognition in different domains, particularly in medical area, have shown promising results in terms of accuracy and execution time for detecting anatomical structures on common imaging modalities. Thanks to advancements in computer hardware (graphical processing unit and tensor processing unit), it is now easier to implement highly complex deep learning and reinforcement learning computing systems. The progress in AI and computational power can reduce the analysis time for modern ICs from months down to only a few hours. The object detection and segmentation/recognition computations for IC images share some similarities with medical images (in terms of having benign and malignant data). The features from the IC images can help in detection and analysis hardware Trojan attacks. However, there are unique challenges for feature detection and extraction in IC images obtained from scanning electron microscopes for usage in hardware security applications that should be carefully studied and addressed. 

In order to make contributions in overcoming the discussed issues, we propose Explainable Vision System for Hardware Testing and Assurance (EVHA) that is a low-cost, accurate, and fast system to detect the smallest possible change to a design. We will engage the state-of-the-art techniques from various fields including, but not limited to hardware security, circuit design and testing, data science, computational imaging, artificial intelligence, and computer vision to develop physical inspection and assurance systems fit for the existing and the emerging failures and threats in the respective industry. The inputs to EVHA are IC SEM images and the output is determination of IC status in terms of having any defect and/or hardware Trojan through addition, deletion, or change in the design cells at the gate/cell-level.

The main part of EVHA utilizes advanced object detection and segmentation/recognition techniques to localize and validate images upon collection by microscope. EVHA needs to be trained on a trusted dataset that includes all types of standard cells on the chip. Such data are usually not available for training and synthetic images for training should be created using generative adversarial networks \cite{du2020wafer}. This system requires a reasonable number of original and synthetic SEM images from each cell type to extract their features and train itself. The system should be able to understand geometrical details and process variations belonging to the same design. This process needs to be done only once for each technology node and chip manufacturer. 

The innovative concept of golden gate/circuit (GGC) introduced here is an integral part of EVHA where only unused spaces on the chip are employed to design a circuit that includes all types of standard cells while they are fully testable. Since the circuit is designed separately, each cell can be logically, but exhaustively, tested in a very short amount of time (in seconds) in a self-test approach with no area overhead added to the original design \cite{xiao2014novel}. Once cells in GGC are verified logically, their physical structure will be imaged and used as an input for training. In the next phase, all other cells on the chip will be authenticated by comparing their physical structure with golden gates. The framework of EVHA determines the exact location of features for comparison with the GDSII file using combined information from the SEM stage coordinates and the extracted image features and compensates for image noise or other imperfections. 

This paper provides an overview of tasks included in the EVHA framework: Intelligent Microscopy for Imaging/Delayering (Task 1); Explainable Block/Object Detection and Recognition of IC SEM Image (Task 2); Golden Gate/Circuits Design and Fabrication (Task 3); and Validation and Security Assessment (Task 4). In summary, the overall objectives of EVHA are stated as: (a) ensuring a 4mm×4mm 22nm IC can be scanned in less than a day; this would be an order-of-magnitude faster compared to existing technologies \cite{vashistha2018trojan}; (b) providing very high-confidence detection of Trojans as long as they have a footprint on the chip layout (note interconnect reliability type Trojans are not supported by EVHA); (c) developing intelligent approach to recognize all design cells (genuine and malicious) with perfect accuracy, considering the manufacturing aspects such as process variations; (d) automating the detection process to minimize human subject matter expert (SME) involvement and reducing the overall cost of trust verification; and (e) ensuring attack resiliency against intelligent adversaries. Meeting these objectives will require an interdisciplinary collaboration among multiple fields.

The rest of this review paper is organized as follows: We discuss the related works in Section \ref{sec:related_works}. A brief overview of the system tasks is provided in Section \ref{sec:system_overview}. In Section \ref{sec:image_analysis}, the Explainable Block/Object Detection and Recognition task is discussed in more detail. Section \ref{sec:results-future_plan} presents description of experiments, achieved results, and their evaluations. Our plans for improvement and completion of system development are given in Section \ref{sec:future-work}. The paper is concluded in Section \ref{sec:conclusion}.

\section{Related Works}\label{sec:related_works}

In this section, the related published works in the area of hardware security based on the proposed attacks and defenses are discussed.

\subsection{Hardware Security - Attacks}
As IC production operations are increasingly outsourced to external foundries to reduce manufacturing costs, there has also been a much larger question of trust between design companies and the foundries, as there is the potential for introducing hardware Trojans during this process \cite{vashistha2018, varshney2020}. These Trojans consist of malicious changes performed on an IC's design and are inserted during manufacturing in order to implement a variety of attacks. Such attacks include sending information to outside sources, allowing an outside source to take control, or even disabling the functionality of a device using the infected IC. Obviously, hardware Trojans pose a potential security threat to their users, which can range from civilian to military applications, where security is absolutely essential for proper functionality. This potential risk has led to an interest in a rapid and accurate hardware Trojan detection system, which ideally would be able to identify the presence of a Trojan before an IC is put into use and can pose a greater risk.

\section{System Overview}\label{sec:system_overview}

\begin{figure}[h!]
    \centering
    \includegraphics[width=\textwidth,keepaspectratio]{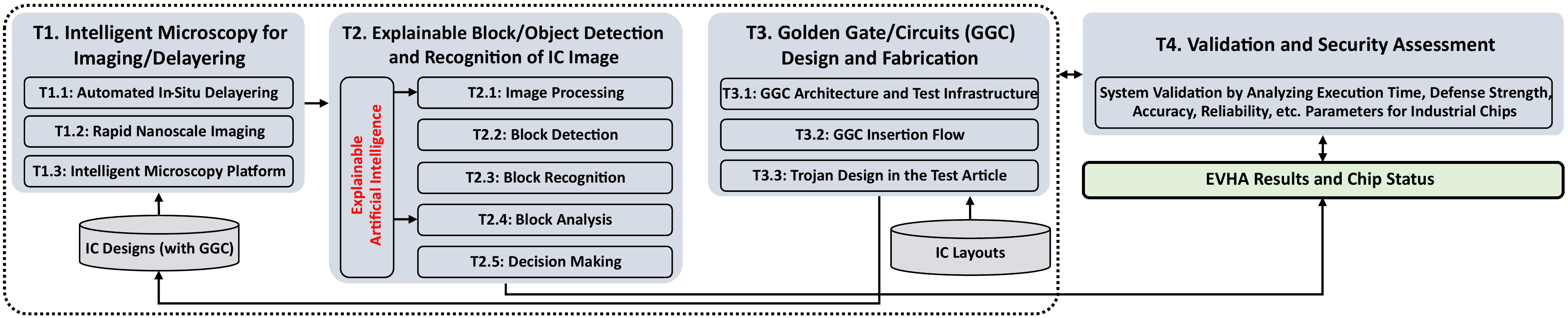}
    \caption{The EVHA framework and its computing tasks.}
    \label{fig:system_tasks}
\end{figure}

We study the four major tasks within the framework of EVHA in this section. In the first task, an auto-delayering tool removes the rest of the silicon, once a backside-thinned IC is loaded into a SEM stage, to expose the doping regions using Monte Carlo simulation for estimation of the remaining thickness of silicon. To the best of our knowledge, it is the first of its kind to perform real-time thickness measurements with no end-point detection probe. Images are intelligently acquired, processed, and analyzed with the GDSII file in consideration. Task 2 focuses on development of novel image analysis algorithms and deep learning architectures to compare SEM images with the original layout, detect Trojans, and assess the results with the highest confidence level. The task also includes an enhancement to the data augmentation and data recognition computations using explainable AI for improving their performance, reliability, and safety. The third task serves in increasing confidence according to which our innovative on-chip circuit with golden reference gates performs self-reference with the chip under imaging. GGC will ensure the obtained training data represents the process variations and structure/features from the chip itself. In Task 4, we validate the EVHA framework by implementing it on industrial-strength test chips fabricated at a 28nm technology node (with golden gates and several stealthy Trojans) while analyzing different attack scenarios against the defense system. Figure \ref{fig:system_tasks} shows the EVHA framework and its described tasks. These tasks are presented in more detail in the rest of this section.

\subsection{Task 1: Sample Preparation and Intelligent Microscopy}

In this subsection, the process of preparing chips based on reducing silicon thickness as well as the process of acquiring IC SEM images are explained.

\subsubsection{Sample Preparation: Backside Polishing of Silicon} \label{sec:Sprep}

Imaging in EVHA requires capturing the active layers of transistors from thick silicon dies sliced from a thick fabricated wafer. Silicon wafers are thick due to a thick substrate, which is typically in the range of hundreds of $um$. The active regions, metal layers, and other device layers are typically in the range of $nm$ thickness. Sample preparation of the silicon die requires removal of silicon to less than one $um$ of Remaining Silicon Thickness (RST). Due to the constraints from the sample preparation tools, the RST is typically in the range of 1-2 $um$. Hence, a silicon die can be more precisely polished to its active layer from the Front End Of the Line (FEOL), which means the backside of the IC. The silicon substrate can be removed faster by milling silicon die and further the die can be polished to remove RST, very near to the active layer. Milling is comparatively faster than polishing and leaves scratches on the silicon die with RST of $10-20um$. Milling is followed by polishing which is comparatively very slower than milling but restores a shiny silicon surface, which is required for backside nano-imaging.

\subsubsection{Intelligent Microscopy: Silicon Plasma Delayering and SEM Imaging of Active Layers}\label{sec:Intelligent_Microscopy}

\begin{figure}[H]
    \centering
    \includegraphics[width=\textwidth,keepaspectratio]{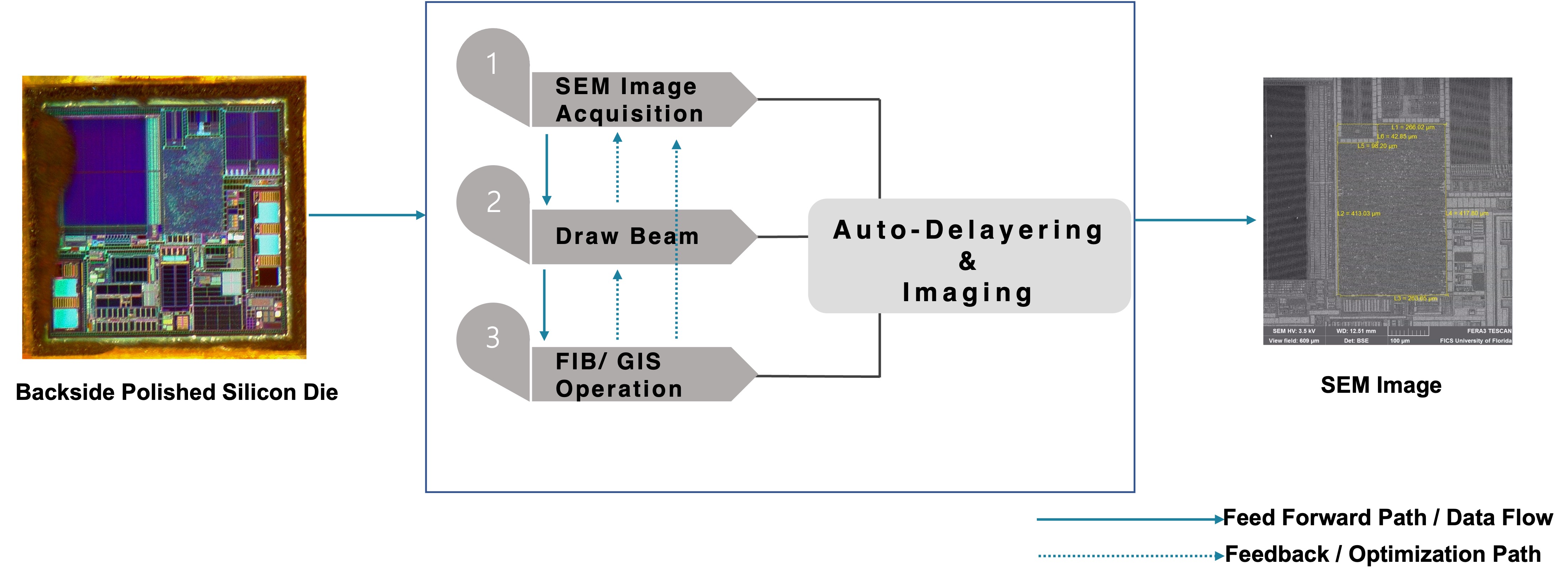}
    \caption{The workflow for auto-delayering and auto-imaging for data acquisition.}
    \label{fig:intel_micro}
\end{figure}

The objective of dual beam nano-level scanning electron microscopy is to capture active regions of the transistor from the backside polished IC or silicon die. After backside silicon removal, the remaining silicon is still opaque to the SEM beam. When imaging a large sample area using high-resolution scanning electron microscopy, the amount of data generated can be overwhelming and the acquisition time can be prohibitive. Furthermore, the capture speed (often referred to as Dwelling Time or DT) can be lengthened in order to ensure higher quality images. Fortunately, an experimenter interested in active layer imaging will often find that only a portion of the data collected is valuable. The image acquisition can then be re-organized into two sequential scans: (1) a large-area scan with resolution adjusted to allow reliable detection of the targets of interest (i.e., primary scan); and (2) a higher-resolution scan with fields of views selectively centered on the targets of interest (i.e., secondary scan). Researchers have named this feedback driven microscopic approach as Intelligent Microscopy. The workflow of this task is displayed in Figure \ref{fig:intel_micro}.

\subsection{Task 2: Explainable Block/Object Detection and Recognition of IC SEM Image}

\begin{figure}[h!]   
\centering
\subfloat[An IC SEM Image. $R_k$ represents a row of cells.]{\includegraphics[width=0.45\textwidth, keepaspectratio]{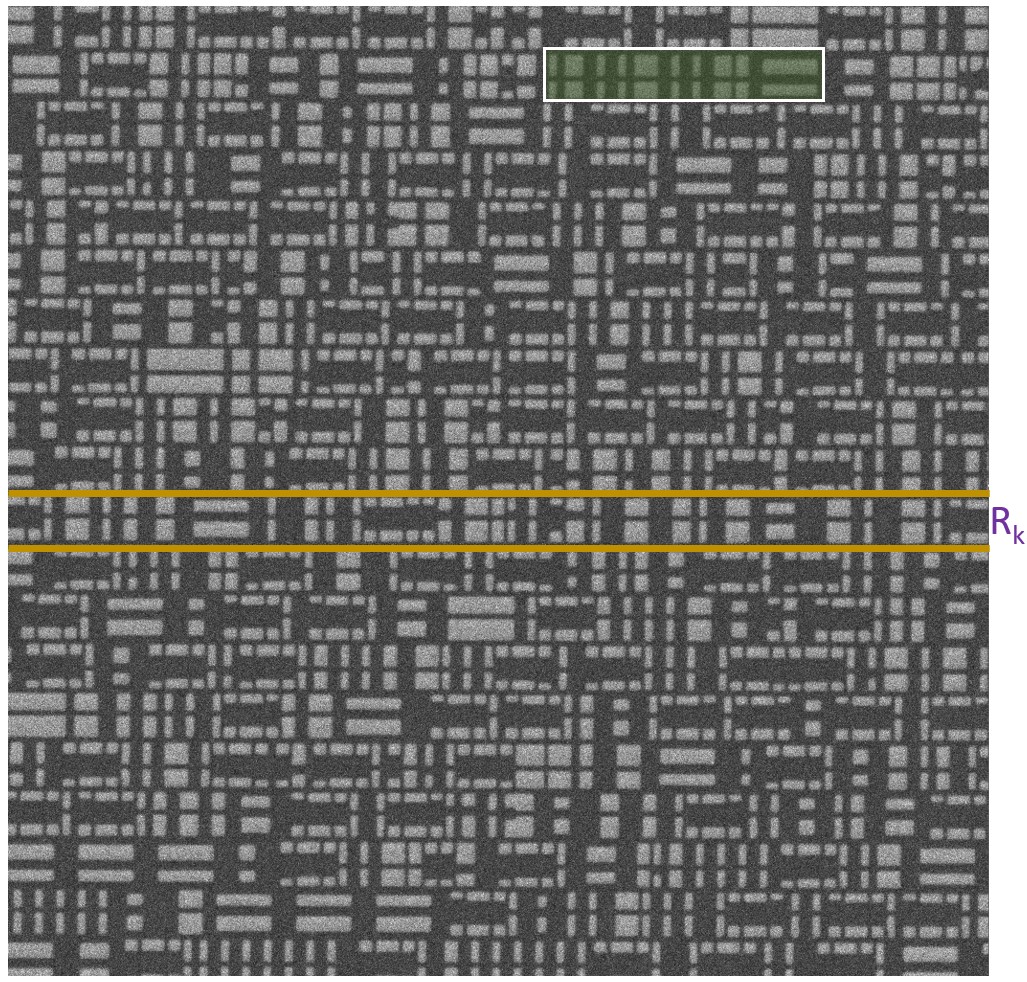}\label{fig:sem_image}}
\hspace{\fill}
\subfloat[The highlighted (green) portion of \ref{fig:sem_image}. Single (blue) and composite (red) cells are marked.]{\includegraphics[width=0.45\textwidth, keepaspectratio]{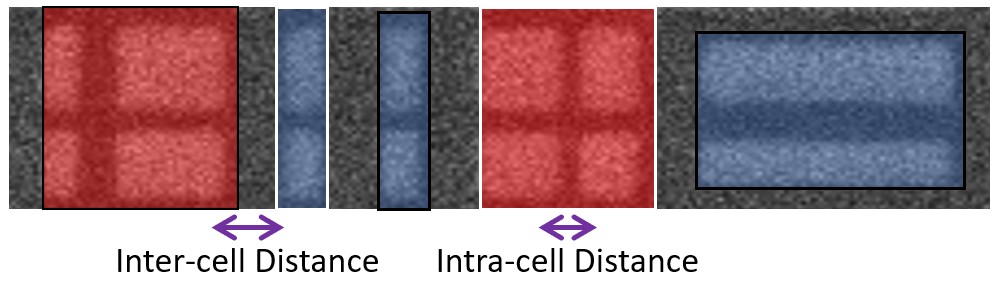}\label{fig:cells}}
\caption{An IC SEM image and its cell structure.}
\label{fig:sem_image_cell}
\end{figure}

The overall objective of this task is extraction of all of the cells available in an IC SEM image in order to analyze them for detection of any possible malicious modifications. The IC SEM images (1024 $\times$ 1024 $\times$ 1) acquired in Section \ref{sec:Intelligent_Microscopy} are passed as the input to the \textbf{"Image Processing"} unit. The computations in this unit are normalization/scaling, denoising, enhancement, and binarization. The type, physical characteristics, and arrangement of cells in the IC SEM image are shown in Figure \ref{fig:sem_image_cell}. 
 
\begin{figure}[h!]
    \centering
    \includegraphics[width=\textwidth,keepaspectratio]{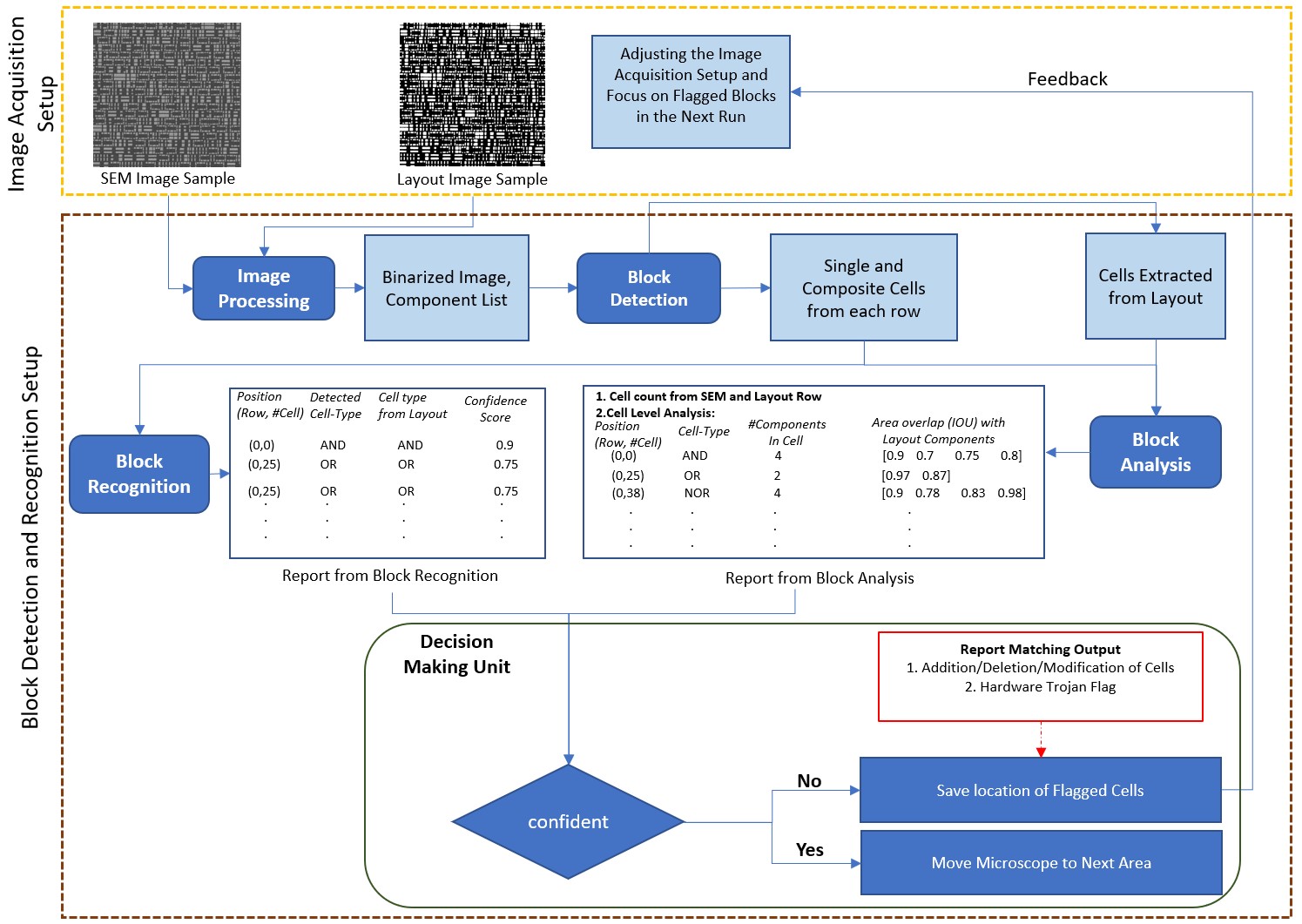}
    \caption{The general view of Task 2: The inputs to the task are the SEM and layout images of IC. The "Image Processing" unit performs normalization/scaling, denoising, enhancement, and binarization on the data. Next, "Block Detection" executes connected component detection on them in order to extract their available cells. The extracted cells are sent to "Block Recognition" and "Block Analysis" units simultaneously. These units evaluate the received data based on the image features and the cell attributes (e.g., the number of cells) respectively. The outputs from them are provided to the "Decision Making" unit in order to determine the chip status. If there is no confidence in the correctness and quality of results, then the "Image Acquisition Setup" is informed and the respective processes are adjusted.}
    \label{fig:image_analysis}
\end{figure}

The \textbf{"Block Detection"} unit extracts logic cells from the processed images, according to which the cells can consist of either single primary gate (e.g., an inverter) or multiple primary gates (e.g., composite cells such as flip-flops), see Figure \ref{fig:cells}. According to the unit computations, the boundaries between consecutive cells (both single and composite) are determined and then the cells are extracted from the entire image. The extracted cells are classified in the \textbf{"Block Recognition"} unit to determine their registration and identify their type. Besides, anomaly level in the images are also measured here. On the other side, a matching analysis is executed by the \textbf{"Block Analysis"} unit on the extracted cells and the Design Exchange Format (DEF) file of the corresponding chip to evaluate their status. The recognition output along with the analysis output are given to the \textbf{"Decision Making"} in order to determine the presence of any defect and/or malicious entity inside the chip under examination. An general view of Task 2 is shown in Figure \ref{fig:image_analysis}

\subsection{Task 3: Golden Gate/Circuits Design and Fabrication}

\begin{figure}[h!]
    \centering
    \includegraphics[width=0.5\textwidth,keepaspectratio]{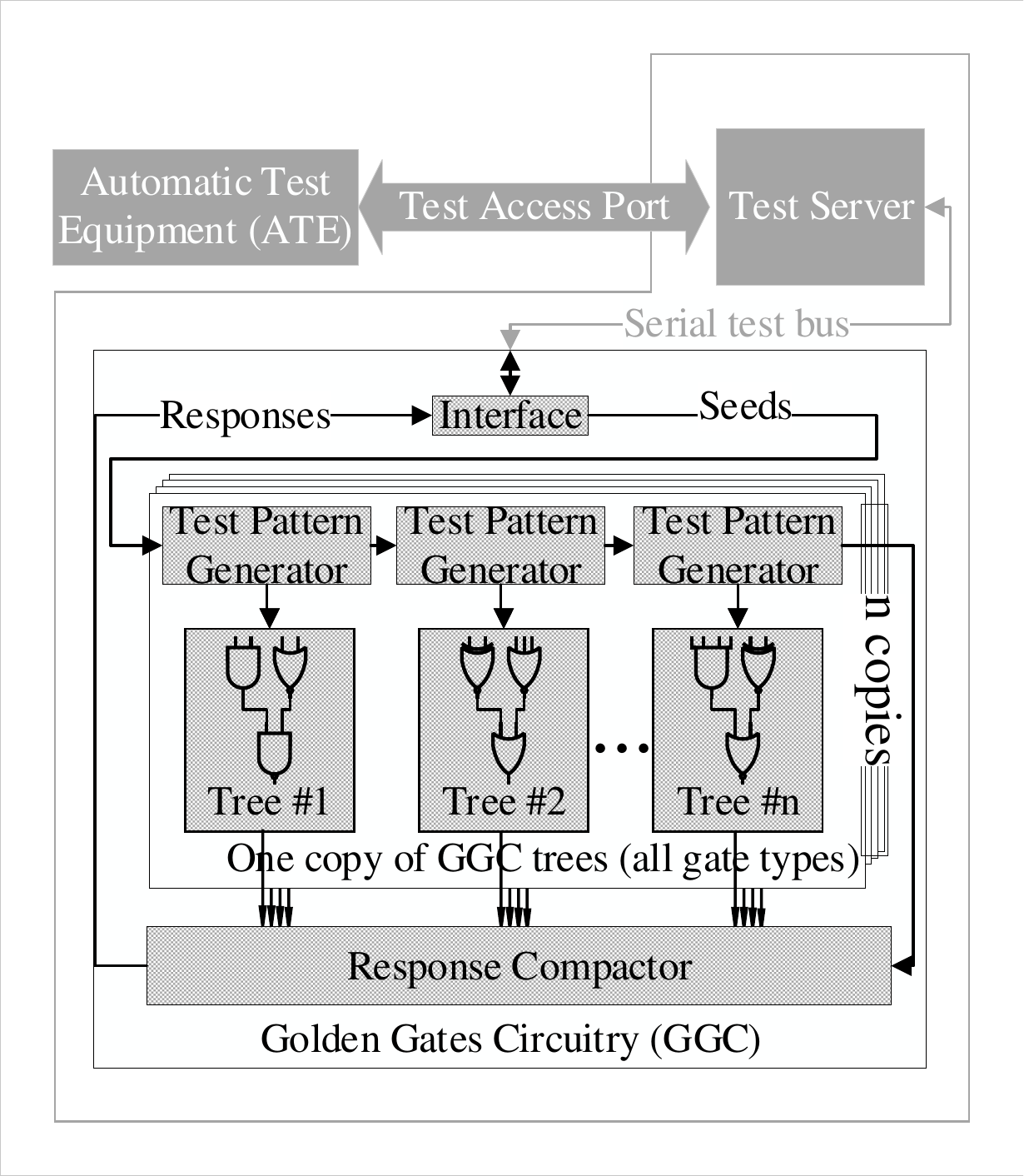}
    \caption{The architecture of a typical GGC in an on-chip testing structure.}
    \label{fig:ggc_arch}
\end{figure}

Due to the lack/insufficiency of reference samples for "training" as well as the need for all samples acquired from the chip(s) belonging to the unknown/untrusted foundry for "testing", we insert golden gate/circuit (which includes the same types of cells as the ones in the original circuit) into the unused space(s) of the original design. This component interacts with a test server through a serial test bus. The test server communicates with an automatic test equipment via a bidirectional test access port. The interface of GGC receives certain data from the test server and it sends corresponding seeds to the test pattern generators. The generators produce data patterns and forward them to the GGC trees. A response compactor collects the information from trees and provides the corresponding responses to the GGC interface that delivers them to the test server and the automatic test equipment. The illustration of described computations is shown in Figure \ref{fig:ggc_arch} \cite{shi2019golden, vashistha2021detecting}.

We also need to have example infected versions of the integrated circuit for examining the assurance strength of EVHA. In this regard, the benchmarks from the "Trust-Hub" repository, specifically those that affect the front-end-of-line features of the design, are employed in which most of the known and hard-to-detect hardware Trojans are included. In addition, design-specific hardware Trojans are designed for measuring EVHA ability in visual assessment of the IC SEM images, with considering a wide range of Kilovolt (KV) levels and different physical profiles. The Trojan-free/golden layout along with the Trojan-inserted/infected layouts are fabricated to validate functionality of the system and assess its security strength.

\subsection{Task 4: Validation and Security Assessment}

Once EVHA is fully developed, it is tested on the fabricated golden and infected chips in order to check its performance, efficiency, speed, and defense potency. There can be two attacks to circumvent the EVHA operations: (a) attacking the GGC authentication by modifying the GGC trees with the purpose of altering the responses. The attack can be properly detected by the computations of Task 2 due to the changes applied on the design cells. (b) insertion of a hardware Trojan with very small overhead that does not cause a noticeable change on the design footprint, leading to inability of system in detecting the malicious modifications. This attack can be prevented through considering redundant logic in the design or creating a power-to-ground short fault in the IC operations. All these experimental processes are run using our in-house devices, instruments, equipment, tools, and software.
\section{ Block Detection and Recognition of IC SEM Images}\label{sec:image_analysis}

A brief overview of the Image Processing, Block Detection, Block Recognition, Block Analysis, and Decision Making units from the second task of EVHA framework are provided in the following.

\subsection{Image Processing Unit} \label{subsec:image_pre-processing}
The defined processes for this unit to be applied on the IC SEM images are stated as normalization/scaling, denoising, enhancement, and binarization. According to our established methodology in \cite{synthetic-sem-image}, the denoising and the binarization computations are described as: (a) \textbf{Denoising:} The images are denoised by the non-local means method. The parameter set-up for the task is followed as described in \cite{non-local-means}, and a denoised image $I_D$ is obtained, refer to Figure \ref{fig:denoised}. (b) \textbf{Binarization:} A binarized Image $I_B$ is obtained by applying global thresholding on the image to convert all pixels to pure white or pure black based on a certain threshold. \cite{Otsu} on $I_D$, refer to Figure \ref{fig:binarized}.\\

\begin{figure}[h!]
    \centering
    \includegraphics[width=\textwidth,keepaspectratio]{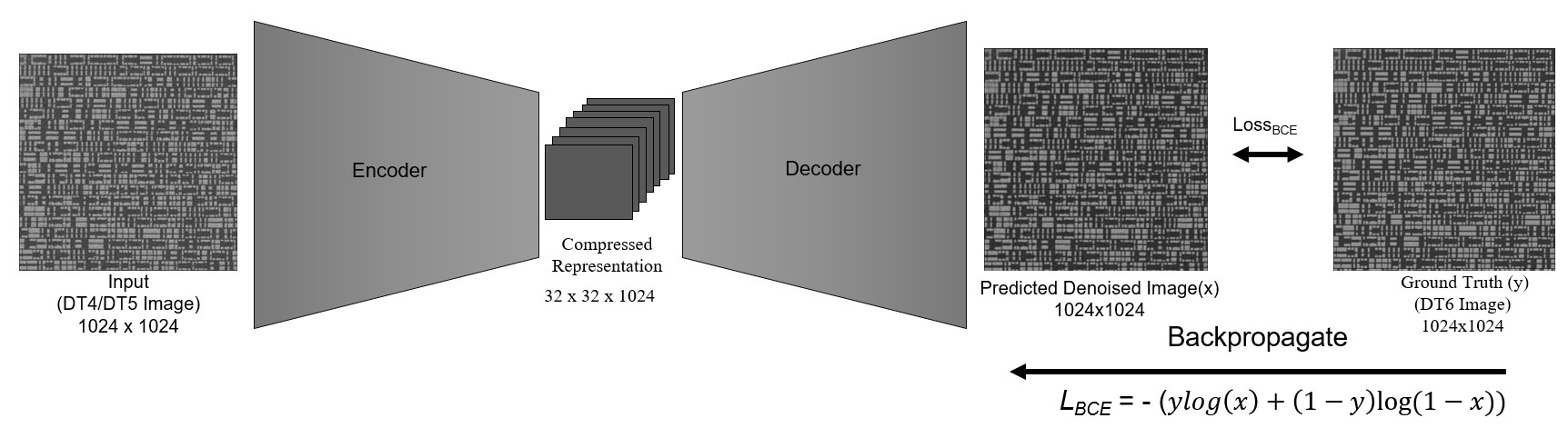}
    \caption{The denoising process of IC SEM images.}
    \label{fig:denoising-system}
\end{figure}

\textbf{\emph{Improved Denoising}.} Images captured with a lower dwelling time (DT) contains higher level of noise. For the successful execution of later computational unit like cell separation (discussed in \ref{subsec:cell-extraction}, a function was learned to denoise higher noisy images by mapping those to corresponding lower noisy version (captured with higher dwelling time). We followed the Noise2Noise \cite{DBLP:journals/corr/abs-1803-04189} setup to execute this denoising task. An encoder maps noisy input images (DT4/DT5 SEM images) into a compressed representation. Using the representation, a decoder generates cleaner images (DT6 SEM images that are less noisy). We have used DT4/DT5 SEM images as the input data and corresponding DT6 SEM image as the ground-truth data for this algorithm. The system architecture of this algorithm is shown in Figure \ref{fig:denoising-system}.

\textbf{\emph{Synthetic Image Generation}.} Due to the necessity of possessing a dataset with diverse and sufficient image samples, another computing process is specified for this unit. In order to achieve a high-performance recognition and evaluation of cell SEM images, it is required to have enough number of diverse samples to perform a strong and comprehensive training for the trainable elements in the recognition and the analysis units. Creating this form of dataset using the existing instruments solely is extremely costly, labor-intensive, and difficult to accomplish. This problem is tackled through applying our methodology for synthetic data generation \cite{synthetic-sem-image} on the original cell SEM images. We have employed Mode-Seeking Generative Adversarial Network (MSGAN) \cite{mao2019mode} for the process.

\subsection{Block Detection Unit} \label{subsec:cell-extraction}
\begin{figure}[h!]   
\centering
\subfloat[A DT6 IC SEM Image.]{\includegraphics[width=0.3\textwidth, keepaspectratio]{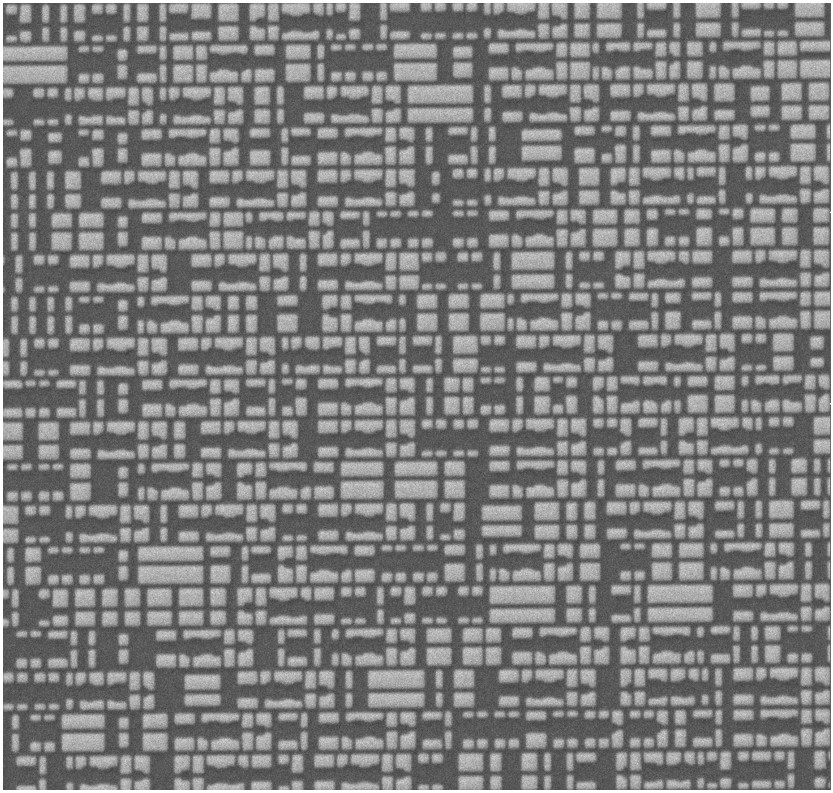}\label{fig:sem_dt5}}%
\hfill
\subfloat[Denoised Version.]{\includegraphics[width=0.3\textwidth, keepaspectratio]{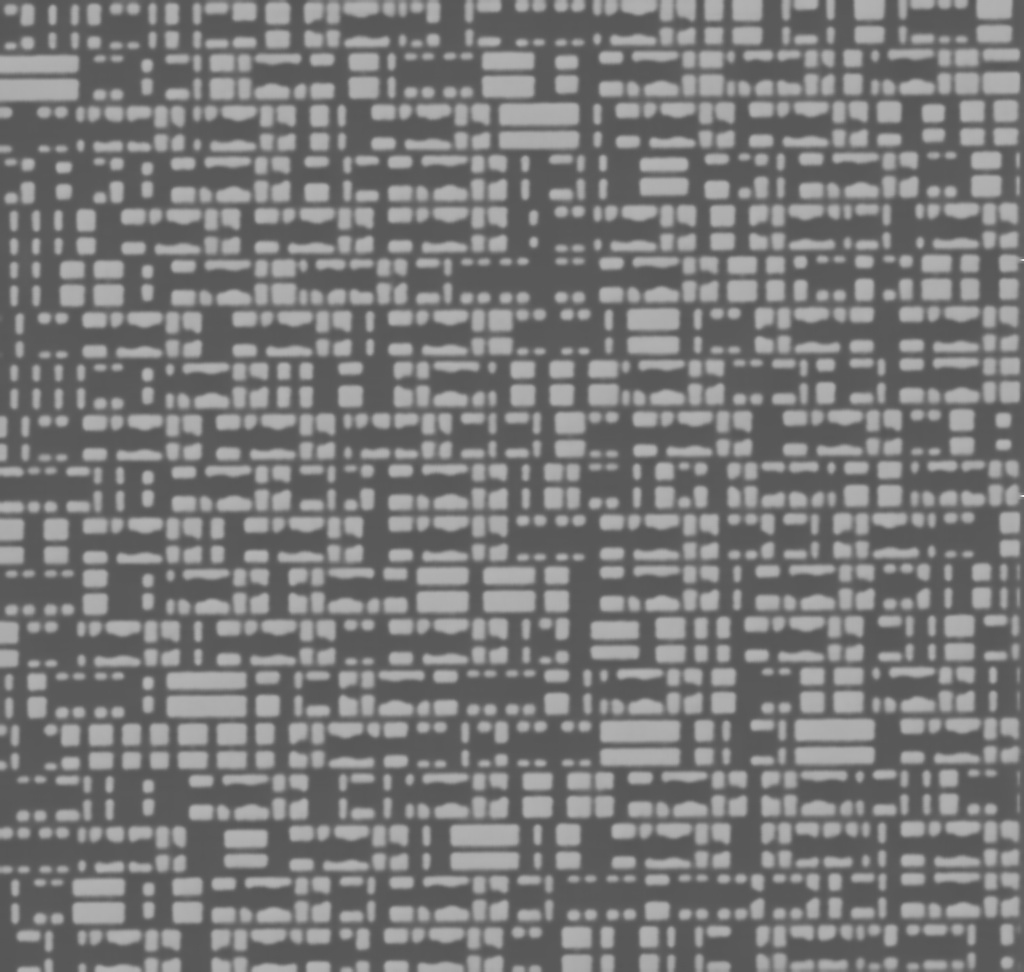}\label{fig:denoised}}%
\hfill
\subfloat[Binarized Version.]{\includegraphics[width=0.3\textwidth, keepaspectratio]{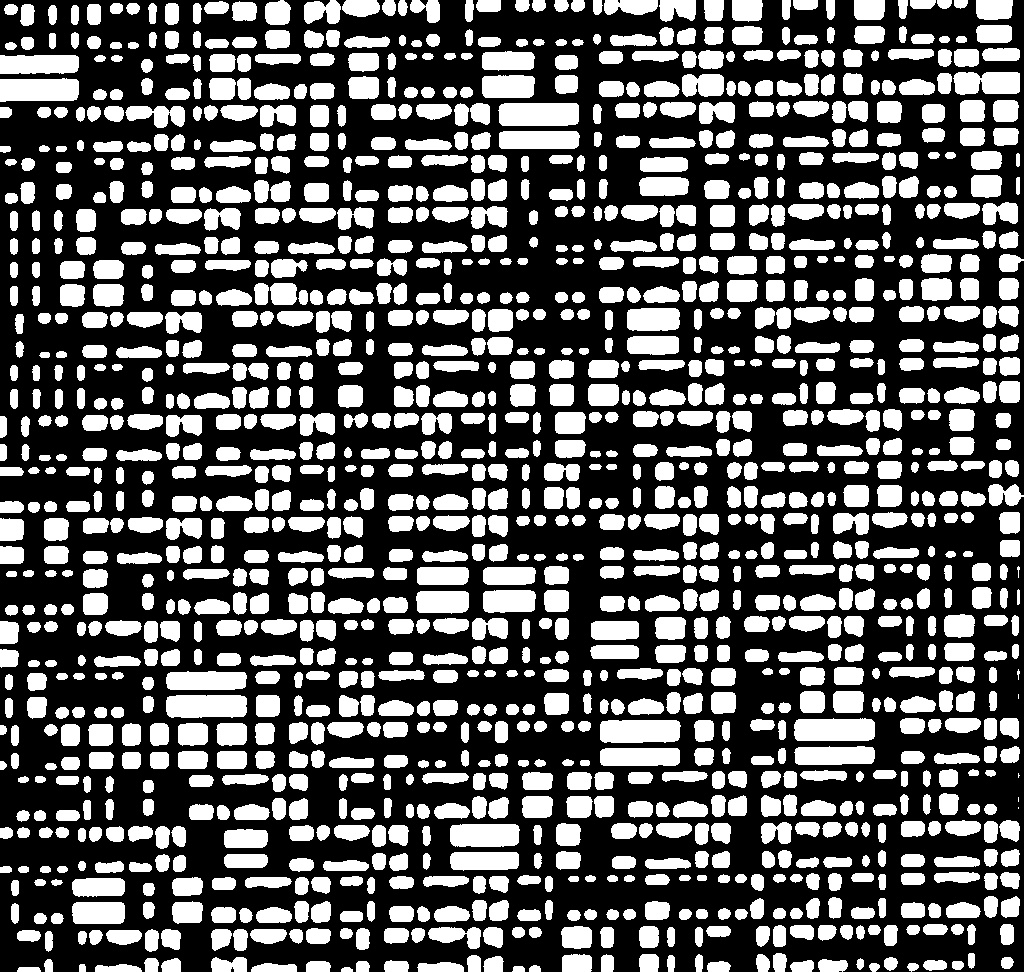}\label{fig:binarized}}
\newline
\subfloat[Image cell components.]{\includegraphics[width=0.3\textwidth, keepaspectratio]{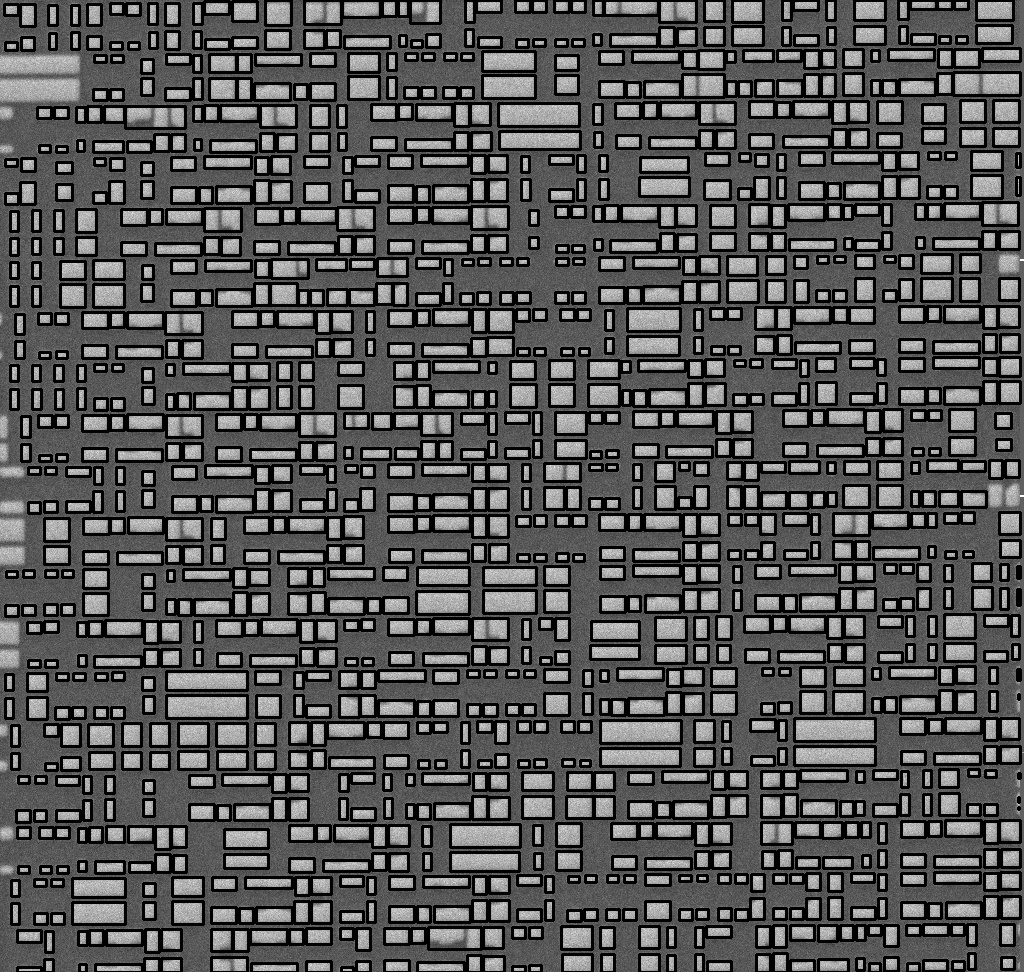}\label{fig:connected_components}}%
\hfill
\subfloat[Identified component.]{\includegraphics[width=0.3\textwidth,keepaspectratio]{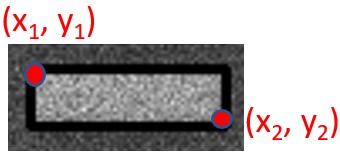}\label{fig:component}}%
\hfill
\subfloat[Eight-connectivity.]{\includegraphics[width=0.3\textwidth, keepaspectratio]{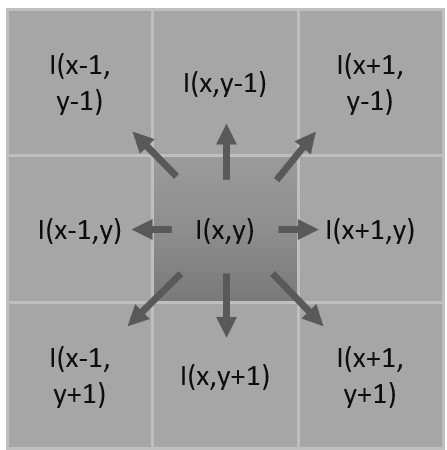}\label{fig:eight-connectivity}}
\caption{The computations of image processing and block detection units.}
\label{fig:pre-processing}
\end{figure}

In this unit, the connected components of IC SEM images are identified and then the cells are extracted from the images. The process of \textbf{Component Identification} is described as: the foreground pixels of an image (which represent white color, i.e., $I(x,y)=255$) are grouped together by scanning $I_B$ from top to bottom and left to right. Eight-connectivity organization of pixels is considered for grouping of pixels with the same intensity value, shown in Figure \ref{fig:eight-connectivity}. A set of distinctively labelled connected components, defined as $C = {c_0, c_1, ... ..., c_{T-1}}$, illustrated in Figure \ref{fig:connected_components}, is obtained in which each $c_i$ represents a component with a top-left $(x_{i-1},y_{i-1})$ and a bottom-right $(x_{i+1},y_{i+1})$ point, see Figure \ref{fig:component}. In $C$ definition, $T$ represents the total number of components in the image $I$. All elements of $C$'s can be sorted vertically by one of the image columns, such as $y_1$. In Figure \ref{fig:sem_dt5}, DT5 refers to the dwelling time of five (i.e., 10 $\mu$s/pixel) in SEM imaging.

The \textbf{Cell Extraction} process includes two major steps and is expressed as \cite{synthetic-sem-image}: (a) \textit{Component Listing}, in which components from each row are listed entirely. An illustration of this computation is shown in Figure \ref{fig:sem_image}. (b) \textit{Cell Separation}, in which single and composite cells are separated out from the component list of each row.

\begin{figure}[h!]   
\centering
\subfloat[Applying row decider on IC SEM images: $R_k$ represents a component row (bounded by two yellow lines). White marked components show the beginning of a new row from a vertically sorted component list.]{\includegraphics[width=0.35\textwidth, keepaspectratio]{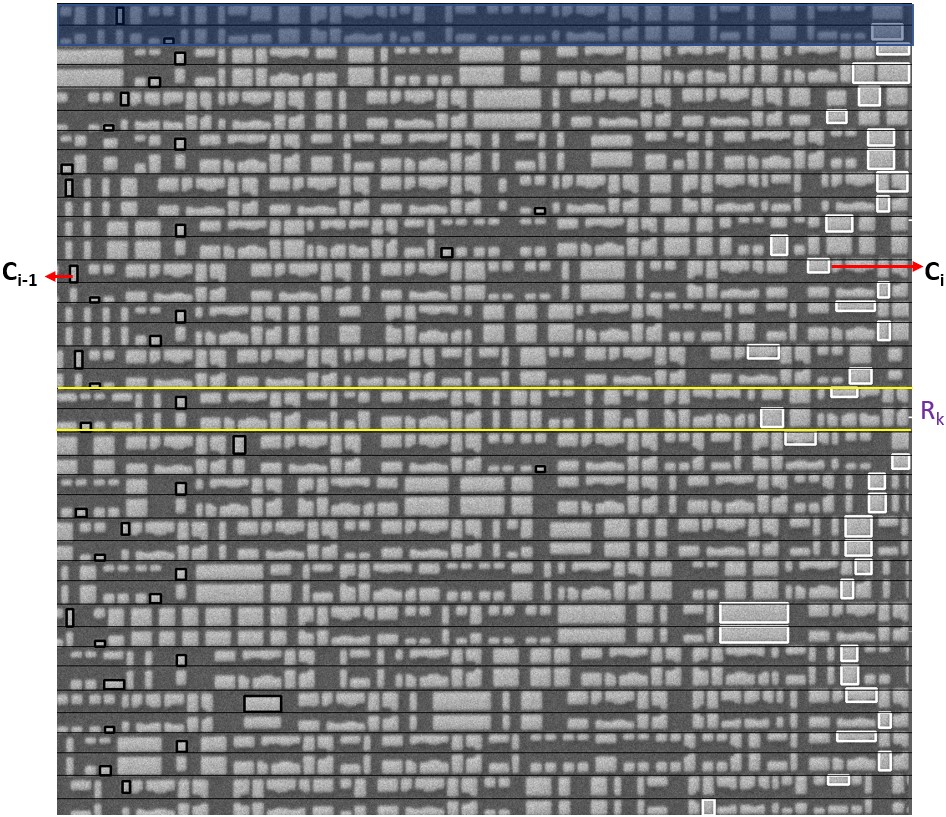}\label{fig:row_separator}}%
\hspace{1em}%
\hfill
\subfloat[Applying cell separator on a marked row of components: (i) shows two parts of a decided row. In (ii)-(iii), consecutive components are merged based on the lateral distances. Cases (iv)-(v) display that the cell boundaries are determined based on the new marked components and then the cells are extracted accordingly.]{\includegraphics[width=0.6\textwidth, keepaspectratio]{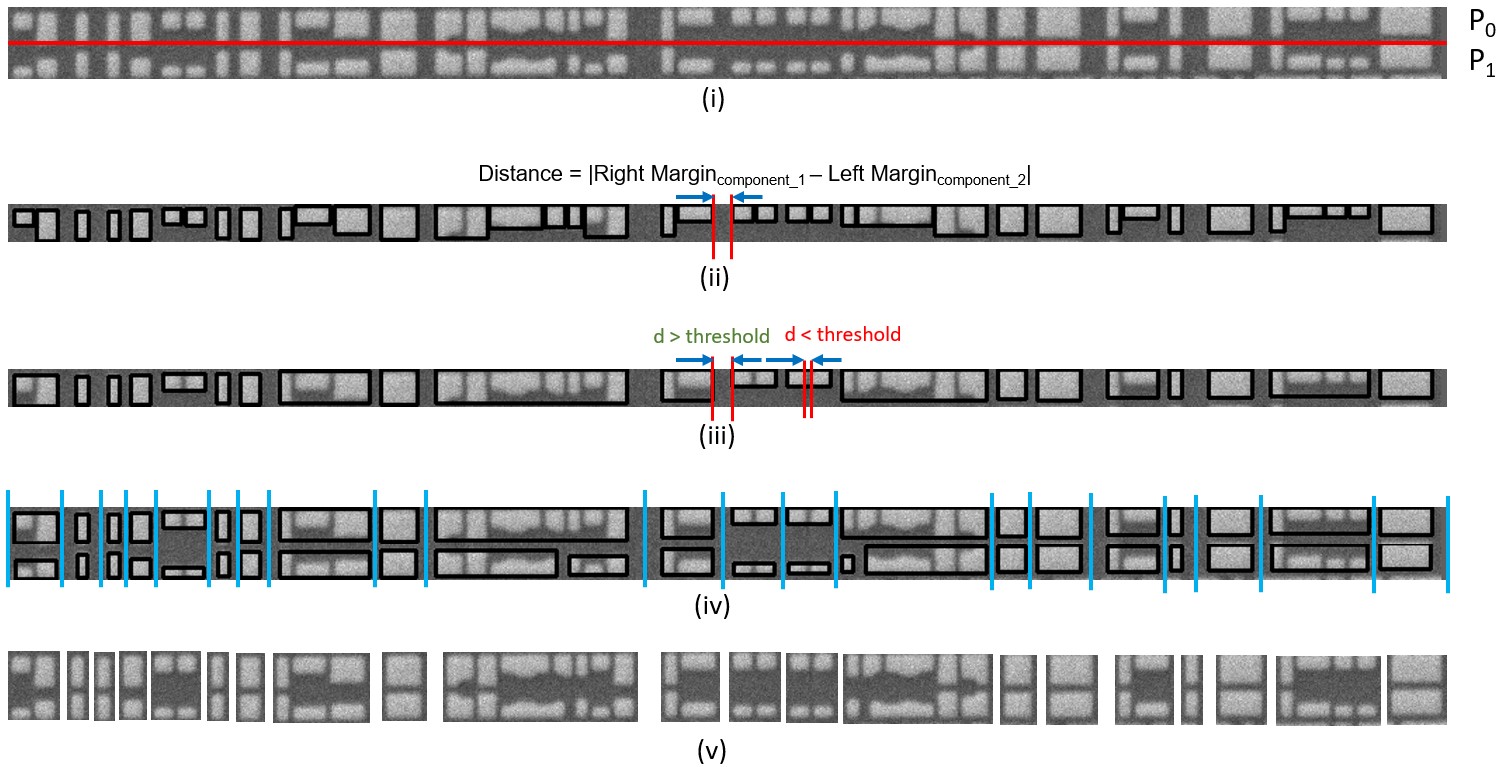}\label{fig:row_processing}}%
\caption{The operations of row decider and cell separator in the image processing unit.}
\label{fig:cell_extraction}
\end{figure}

In the first major step, the beginning of a new component row is marked by making comparison between the y-coordinates of successive components. While traversing the vertically sorted components, a component $C_i$ is assumed to be a part of a new row if it starts from a higher position than its previous components, which means $C_i.y_1 >= C_{i-1}.y_2$. This concept is depicted in Figure \ref{fig:row_separator}. This comparison works based on the assumption that the connected components are calculated accurately and the images are not significantly tilted. The components of each separated row are then sorted according to their top-left coordinates ($x_1$). The formulation of procedure is presented in Algorithm \ref{alg:component_listing}.

\begin{algorithm}[h!]
\caption{The component listing procedure.}\label{alg:component_listing}
\begin{flushleft}
\hspace*{\algorithmicindent} \textbf{Input:} Connected Components, $C[C_0,C_1, ... ..., C_{T-1}]$ where each $C_i = [x_1, y_1, x_2, y_2]$\\
\hspace*{\algorithmicindent} \textbf{Output:} Component list $C_R$ from each row $R_k$
\end{flushleft}
\begin{algorithmic}[1]
\Procedure{Row-WiseComponentList}{}
\State $\textit{$C_R$} \gets List() $
\State $\textit{$Components$} \gets List()$
\For{$i \gets 1$ to $C.length-1$}
    \State {Components.append($C_{i-1}$)}
    \If {$C_i.y_1 \geq C_{i-1}.y_2$}
        \State $C_R.append(Components)$
        \State $Components \gets List()$
    \EndIf
\EndFor
\EndProcedure
\end{algorithmic}
\end{algorithm}

With respect to the second major step, the components of each row are scanned to separate out single and multiple component cells. In most of the cases, each row($R_k$) is composed of two symmetric parts, such as $P_0$ and $P_1$ shown in Figure \ref{fig:row_processing}(i). The separation computations can be performed on either $P_0$ or $P_1$ as shown in Figure \ref{fig:row_processing}. Whether two components belong to the same or different cells is decided based on the distances between the consecutive components, which can be observed in Figure \ref{fig:row_processing}(ii)-(iii). Cell margins are determined according to new components and the cells are extracted according to the new format, as the processes are displayed in Figure \ref{fig:row_processing}(iv)-(v). This procedure is demonstrated mathematically in Algorithm \ref{alg:cell_separation}.

\begin{algorithm}[h!]
\caption{Cell Separation}\label{alg:cell_separation}
\begin{flushleft}
\hspace*{\algorithmicindent}\textbf{Input:} List of Components for each row, $C_R[C_{R_0},C_{R_1}, ... ..., C_{R_{N-1}}]$ where N is the total number of Rows and each $C_{R_i} = [C_0, C_1, C_2, ... ..., C_{m-1}]$; here each $C_i = [x_1, y_1, x_2, y_2]$ and m varies from row-to-row.\\
\hspace*{\algorithmicindent}\textbf{Output:} Merged Component List $C_R$
\end{flushleft}
\begin{algorithmic}[1]
\Procedure{CellSeparation}{}
\State $i \gets \textit{0} $
\State $threshold \gets \textit{P} $
\State $Components \gets List()$
\While{$i \textless C_R.length - 1$}
    \State $C \gets C_{R_i}$ 
    \For{$j \gets 1$ to $C.length - 1$}
        \If {$|C_j.x_1 - C_{j-1}.x_2|_2 < threshold$}
            \State $C_j \gets merge(C_{j-1},C_j)$
            \State $C_{R_{ij}} \gets C_j$
        \EndIf
    \EndFor
    \State $i \gets i + 1 $
\EndWhile
\EndProcedure
\end{algorithmic}
\end{algorithm}

The above computations are not effective in certain image conditions, such as images with high levels noise or severe rotation(s). This methodology serves as a proof of concept and introduces a research direction for comprehensive studies with stronger outcomes. In this regard, our future plan for improving the discussed algorithms and the overall functionality of block detection unit are provided in Section \ref{sec:future-work}.


\subsection{Block Recognition Unit}\label{subsec:blk_recognition}
The block recognition unit works as a filter on what to pass to the rest of the process. The unit receives cells extracted from the SEM image. The task of the unit is two-fold.

\begin{enumerate}
    \item Identify type of the input cells
    \item Filter out anomalous cell image 
\end{enumerate}

\begin{figure}[h!]
    \centering
    \includegraphics[width=\textwidth]{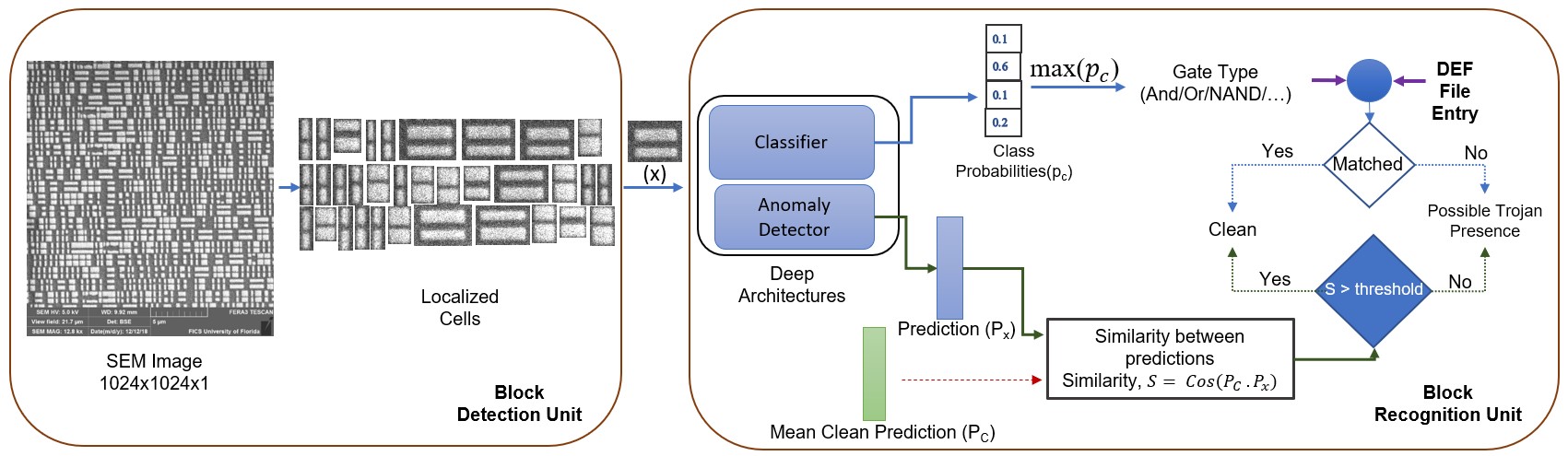}
    \caption{The overall computational flow of block recognition unit.}
    \label{fig:blk_det_recog}
\end{figure}

The classifier of the unit outputs a probability
distribution corresponding to the number of types/classes of IC cells and uses it to determine the appropriate class for a cell (which is specified based on the highest probability in the class vector). The registration of the reference cells is done during the training process using the image format of DEF layout file. The classifier output determines the validity of a cell in terms of the presence of its type in the original design.\\
\hspace*{3mm}On the other hand, the anomaly detection branch determines whether a cell image was present during the registration process or the image of a registered cell is changed sufficiently. If it does, then the cell status is set as "unknown" that means presence of an abnormal entity (malicious or non-malicious) in the design. The overall computational flow of block recognition unit is depicted in Figure \ref{fig:blk_det_recog}. A report containing the class probabilities and anomaly status is generated from the recognition unit and is delivered to the decision-making unit.

\begin{figure}[h!]
\centering
\includegraphics[width=\textwidth]{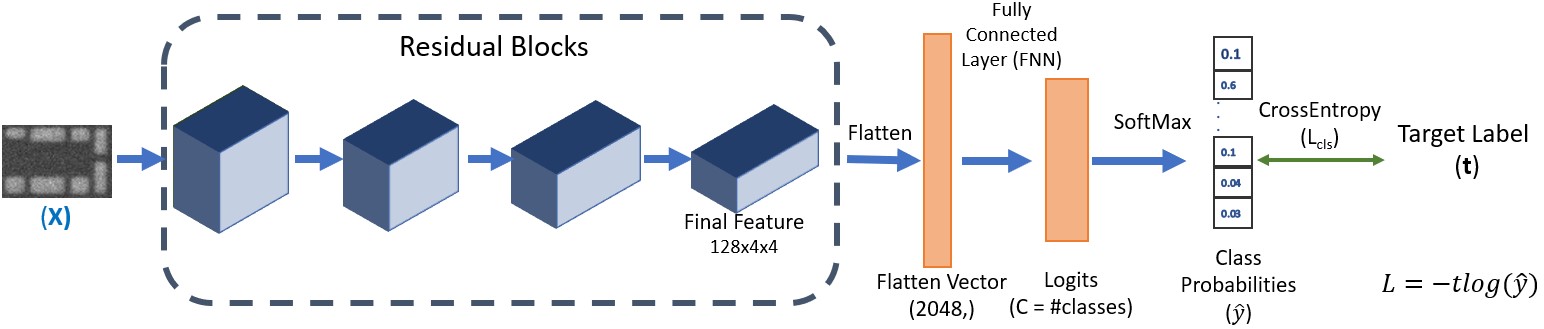}
\caption{Training of the classifier.}
\label{fig:classification}
\end{figure}

\textbf{\emph{Identification of Cells}.} For classification task, unlike previous studies \cite{vashistha2018}, We have adopted a Convolutional Neural Network (CNN)-based architecture to classify the extracted cells, which require large amounts of data for training. The training data is developed through imaging (i.e., original data acquisition) and synthetic data generation as mentioned in \ref{subsec:image_pre-processing}. The extracted and synthesized images of cells, along with their labels, are fed into the classifier for training. Input image($x$) is passed through the classifier and softmax (\cite{softMax}) is applied on the computed logit to get the probability distribution. Finally, cross-entropy loss is applied between predicted probability and available ground-truth label of the corresponding input. The classifier training is depicted in Figure \ref{fig:classification}.

\begin{figure}[h!]
\centering
\includegraphics[width=\textwidth]{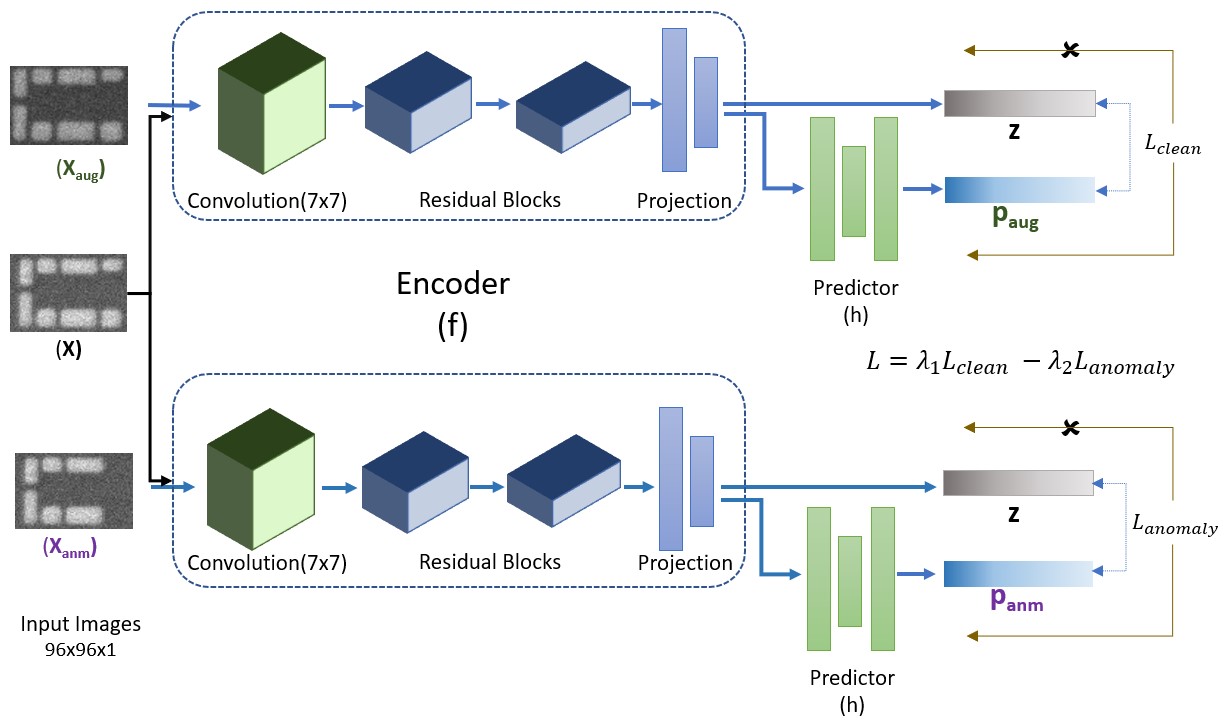}
\caption{Training of the anomaly detector is depicted. Original view of the image ($X$) and its augmented ($X_{aug}$) and anomalous ($X_{anm}$) view are passed through the network. The similarity is calculated between the projection of one view (like $z$) and prediction output of another view (like $p_{aug}$). Stopgrad is applied for better convergence of the task ($\times$ marked path in the figure). The encoder and predictor are shared among the inputs. For better understanding the network is drawn twice here.}
\label{fig:anomaly-detection}
\end{figure}

\textbf{\emph{Anomalous Cell Prediction}.} For our training dataset $D_{t} = \{x_{n}\}_{n=1}^N$, which consists of anomaly free images, the target is to train the system such that it can detect anomalous/trojan cells during test time. We adopted the simple siamese self-supervised learning (SSL) framework \cite{Chen2021ExploringSS} that was originally intended for unsupervised representation learning. Training of the task is depicted in Figure \ref{fig:anomaly-detection}.\\

\hspace*{3mm} Unlike conventional learning system, the loss function runs over pairs of samples during training the system. Our architecture (Figure \ref{fig:anomaly-detection}) takes input as augmented view $x_{aug}$ and anomalous view $x_{anm}$ of the original image $x$ along with the image itself. Input $x$ is transformed by a stochastic data augmentation module to produce the correlated view $x_{aug}$. Anomalous view $x_{anm}$ is produced by randomly cropping a part of input image $x$ and paste it elsewhere on the image as described in \cite{Li2021CutPasteSL}.\\
\hspace*{3mm} These inputs are processed by an encoder $f$ consisting of a backbone (ResNet-18 \cite{He2016DeepRL}) and a projection head as shown in Figure \ref{fig:anomaly-detection}. The prediction head $h$ transforms the output of one view and calculate similarity with other view. Denoting two output vectors $p = h(f(x))$ and $z_{aug} = f(x_{aug})$, the negative cosine similarity is minimized over the training:
\begin{equation}\label{eq:cosine-similarity}
    D(p, z_{aug}) = - \frac{p}{\lVert p \rVert_2} \cdot \frac{z_{aug}}{\lVert z_{aug} \rVert_2}
\end{equation}
\hspace*{3mm} where $\lVert . \rVert$ is $l_2$ norm. The symmetrized loss between $x$ and $x_{aug}$ is written as in Equation \ref{eq:loss-clean-pair}
\begin{equation}\label{eq:loss-clean-pair}
    L_{clean} =\frac{1}{2}D(p, stopgrad(z_{aug})) + \frac{1}{2}D(p_{aug}, stopgrad(z))
\end{equation}
The stopgrad operation kind of simulates the expectation-maximization like algorithm as hypothesised in \cite{Chen2021ExploringSS} and it's an imperative part for better convergence of SSL.
\hspace*{3mm} Similarly, The symmetrized loss between $x$ and $x_{anm}$ can be written as in Equation \ref{eq:loss-anomalous-pair}.
\begin{equation}\label{eq:loss-anomalous-pair}
    L_{anomaly} =\frac{1}{2}D(p, stopgrad(z_{anm})) + \frac{1}{2}D(p_{anm}, stopgrad(z))
\end{equation}
\hspace*{3mm} As the target is to maximize similarity of original view of input $x$ with augmented view $x_{aug}$ while minimizing similarity with anomalous view $x_{anm}$, the final objective function can be written as the following.
\begin{equation}\label{eq:loss-anomalous-pair}
    L = \lambda_1 L_{clean} - \lambda_2 L_{anomaly}
\end{equation}
where $\lambda_1$,$\lambda_2 \in [0,1]$ are hyper-parameters.

During inference, we passed an extracted cell image ($x_{test}$) through the network and calculate the prediction vector $p_{x_{test}}$ . To identify the malicious label, first mean clean prediction vector $p_c$ is determined beforehand using prediction vectors from original training images. Then based on the similarity between $p_{x_{test}}$ and $p_c$, the anomaly level of $x_{test}$ is decided. If the similarity falls below a threshold, then the sample status is set as "malicious" and a possible trojan signal is triggered (see Fig. \ref{fig:blk_det_recog})
\begin{equation}\label{eq:mean_clean_embd_dist}
    S = \cos(p_c, p_{x_{test}})
\end{equation}

\textbf{\emph{Training Details}}. All of the cell SEM images have been distributed into training and validation following a $80/20$ split. The images are then resized to a resolution of 96 $\times$ 96 pixels before being advanced through the network.\\
\hspace*{3mm} For classification task, pre-processing techniques like Gaussian blurring, Gaussian noise addition, rotation, vertical flip etc. are applied to the input image to create augmented pair and to mimic certain real-world scenarios of SEM imaging.\\
\hspace*{3mm} For anomaly detection task, random-crop, horizontal flip, Gaussian blurring and color jittering have been used to produce the augmented view of the original image. As mentioned earlier, for anomalous view $x_{anm}$, we followed the method as described in \cite{Li2021CutPasteSL}. We have experimented with different batch sizes (128, 256, 512) for the training of anomaly detector but in our case, batch size of 128 gives the best outcome. The network is trained only with 
positive pairs ($x$ and $x_{aug}$) for some initial warm-up epoch (15) and later we incorporated the negative pairs ($x$ and $x_{anm}$).\\
\hspace*{3mm} A learning rate of 0.0001  and as optimizer SGD \cite{Ruder2016AnOO} is used throughout the training of both methods.\\
\textbf{\emph{Decision Evaluation}}. To evaluate and visualize the classifier performance, heatmaps of results are generated using the Gradient-weighted Class Activation Mapping (Grad-CAM) algorithm ~\cite{Selvaraju_2017_ICCV}. The heatmap is used to demonstrate the regions activated by the model.\\
\hspace*{3mm} For anomaly detection, the similarity score between mean prediction from clean training images and test image prediction is calculated. A qualitative analysis on the decisions is presented in section \ref{sec:results-future_plan}.

\subsection{Block Analysis Unit} \label{subsec:blk_analysis}
Cells identified as "non-malicious" in \ref{subsec:blk_recognition} are passed to a block analysis unit for further verification by a subtle structure level analysis. Here dopant regions of each cell are compared against corresponding regions in the golden layout image in order to find the dis-similarities. So, the amount of similarities and differences between the pairs of elements (components from SEM image cell and layout image) is used to identify abnormal modifications and determine presence of defect and/or hardware Trojan in the chip under test. Finally, a report is produced depicting the status of examined cells and is provided to the decision-making unit.

\begin{figure}[h!]  
\centering
\subfloat[Counting the number of cells for the SEM and the layout images.]{\includegraphics[width=0.8\textwidth, keepaspectratio]{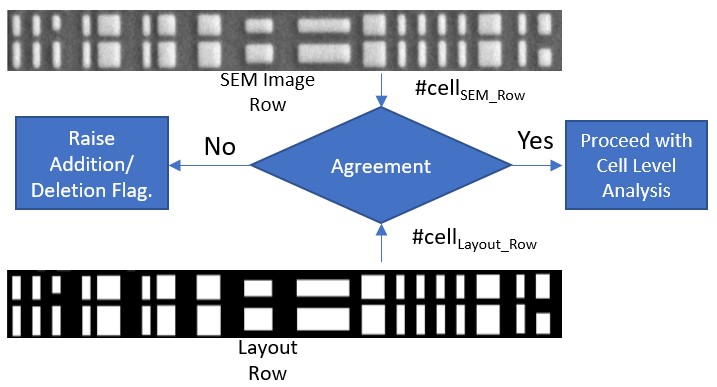}\label{fig:cell_count}}
\newline
\subfloat[Correspondence finding between the polygons of SEM and layout cell images. The cell regions of SEM and layout images are marked in pink and blue colors respectively.]{\includegraphics[width=0.65\textwidth, keepaspectratio]{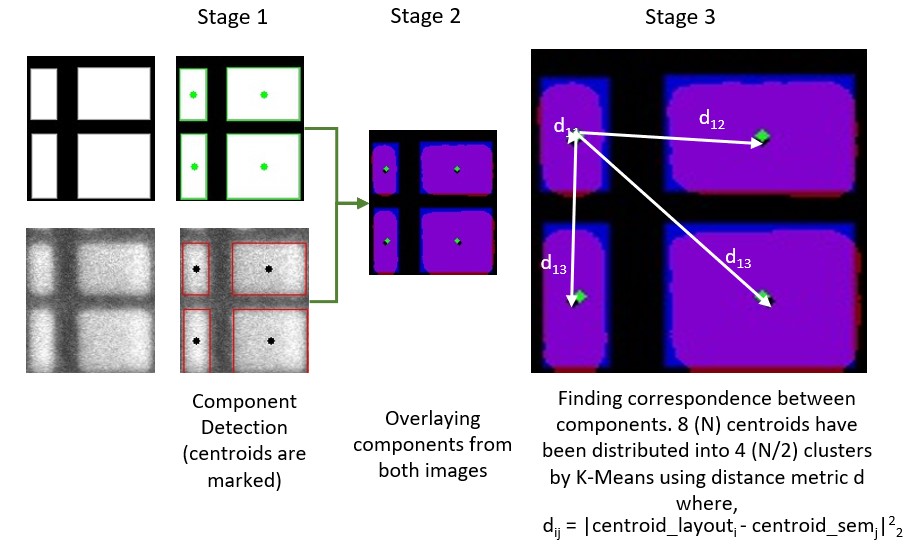}\label{fig:correspondence}}
\caption{The two major steps in the block analysis unit.}
\label{fig:block_analysis}
\end{figure}

Characteristics like number of cells, cell area (i.e., height and width), location, and deformity-level are considered during analysis. Block analysis includes two major steps: (a) counting all the cells, and (b) performing cell-level analysis. cells in every row of the IC SEM image and the golden layout images are enumerated. If there's a mismatch in the number of cells, then a warning flag is raised and the status of the corresponding row is included in the output report. Otherwise, we proceed with cell level analysis. The workflow of this stage is shown in Figure \ref{fig:cell_count}.

In cell level analysis, we closely followed the work described in \cite{9816465} for cell level validation. Components of a cell are calculated for the test and reference layout and then they are compared. First, bounding boxes enclosing dopant regions are determined for both the test and the reference cell. Centroid values are calculated from each box. In order to establish a one-to-one correspondence between the cells, the SEM image (test data) is binarized and laid over the layout image (reference data). Then distances between the bounding box centroids are calculated according to Equation \ref{eq:correspondence_layout-sem}, and K-Means clustering \cite{Jin2010} is applied on the total number of centroids (N) to create $\frac{N}{2}$ clusters, where each cluster corresponds to a layout centroid and its corresponding SEM centroid. In this equation, j=[1,M] and M is the number of centroids in the SEM image cell under consideration. The three stages are illustrated in Figure \ref{fig:correspondence}.

\begin{equation}\label{eq:correspondence_layout-sem}
\begin{split}
    d_{ij} = |Centroid_{Layout_{i}}-Centroid_{SEM_{j}}|_{2}^{2}
\end{split}
\end{equation}

\begin{figure}[h!]
    \centering
    \includegraphics[width=0.3\textwidth, keepaspectratio]{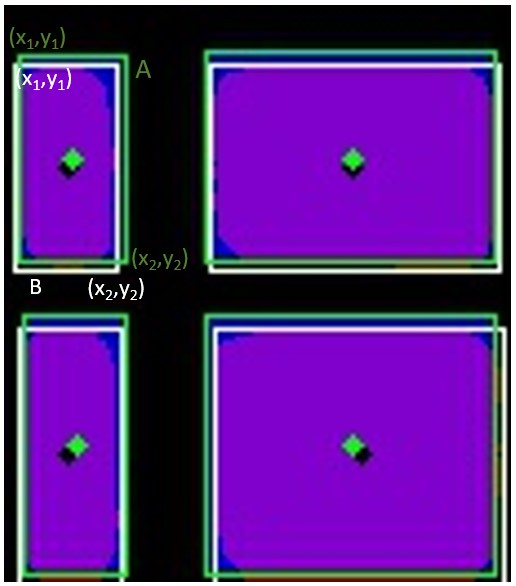}
    \caption{The Intersection Over Union (IOU) accuracy measure applied between the polygon bounding boxes from cell SEM image (A) and its corresponding cell layout image (B).}
    \label{fig:iou}
\end{figure}

In order to measure the deformity-level in the test images, the overlap between test and reference cell components is measured. To measure overlap, we opt to use the Jaccard Similarity Index, also known as Intersection-Over-Union (IOU), which is measured as the ratio of the area of a shared region over the total region (refer to Equation \ref{eq:iou}). The bounding boxes calculated in the matching process earlier are used again for this operation, and are denoted as "A" for the SEM box and "B" for the layout box (see Figure \ref{fig:iou}). For each pair of corresponding boxes, the shared and the total regions are calculated as in Equation \ref{eq:intersection-union}, and the IOU score is calculated according to Equation \ref{eq:iou}. If the IOU is below a certain threshold for any of the boxes within a cell, then the whole cell is marked as abnormal/malicious. All of the measurements are reported to the decision-making unit.

\begin{equation}\label{eq:intersection-union}
\begin{split}
    & LeftMost = max(A.x_1, B.x_1) 
    \\
    & TopMost = max(A.y_1, B.y_1)  
    \\
    & RightMost = min(A.x_2, B.x_2)  
    \\
    & BottomMost = min(A.y_2, B.y_2)  
    \\
    & |A \cap B| = (RightMost - LeftMost) \times (BottomMost - TopMost)
    \\
    & |A \cup B| = Area(A) + Area(B) - |A \cap B| 
\end{split}
\end{equation}

\begin{equation} \label{eq:iou}
    IOU(A,B) = \frac{|A \cap B|}{|A \cup B|}
\end{equation}

\subsection{Decision Making Unit} \label{subsec:decision}

The decision-making unit of EVHA has the duty of determining the IC status in terms of having any defect and/or hardware Trojan through addition, deletion, or change in the design cells at the gate/cell-level. It performs an automated textual analysis on the received reports from the recognition and the analysis units. The parameters in its computations are probability scores for cell images, number of cells, cell area (i.e., height and width), location, and deformity-level. Having reference data for these parameters is extremely critical in order to achieve a comprehensive, reliable, and trustworthy system results.

\section{Results and Discussion}\label{sec:results-future_plan}

We have conducted experiments on each subsystem separately with its specific technical conditions and using limited amount of data (due to in-progress data acquisition and synthetic data generation tasks). With developing a fully diverse and sufficient dataset, our future work will present comprehensive and detailed results, analyses, and interpretations. In the following, the achieved results and related discussions for the discussed computations are provided.

\begin{figure}[h!]   
\centering
\subfloat[Noisy input (DT5 Image)]{\includegraphics[width=0.48\textwidth, keepaspectratio]{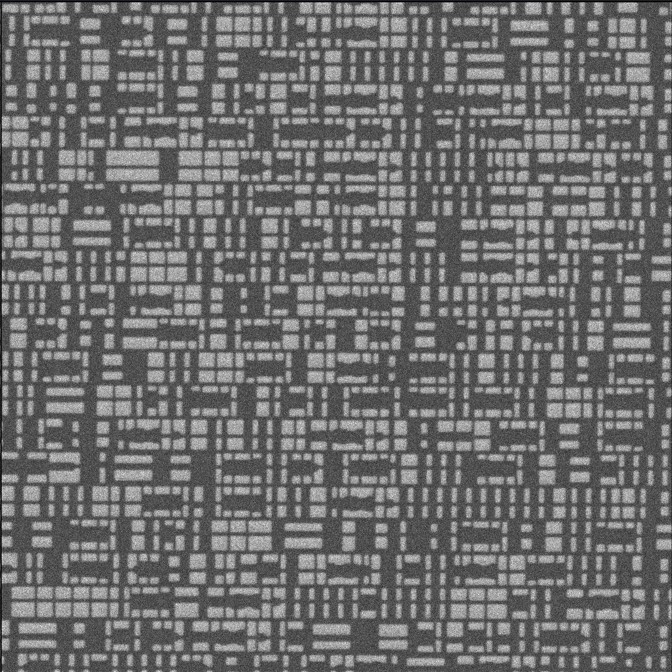}\label{fig:noisy_input}}%
\hfill
\subfloat[Ground truth (DT6 image)]{\includegraphics[width=0.48\textwidth, keepaspectratio]{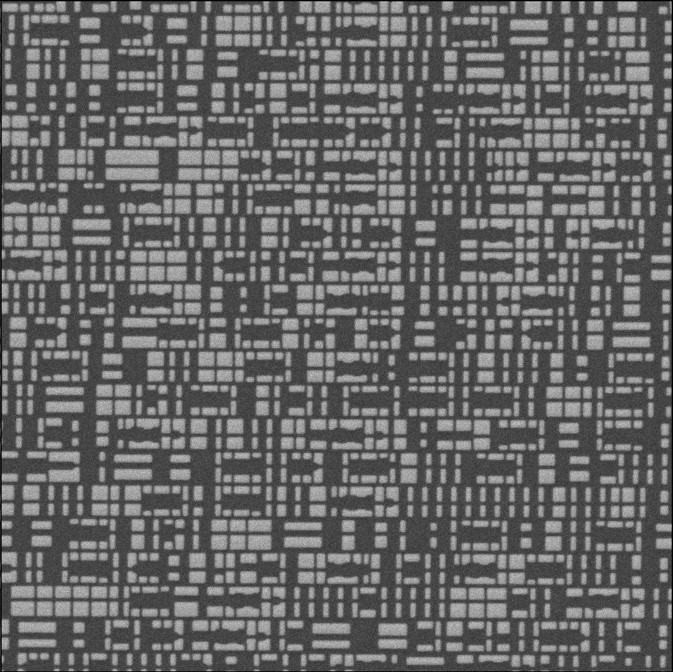}\label{fig:target_image}}%
\newline
\subfloat[Denoised using $L_0$ loss. PSNR: 22.92 dB]{\includegraphics[width=0.48\textwidth, keepaspectratio]{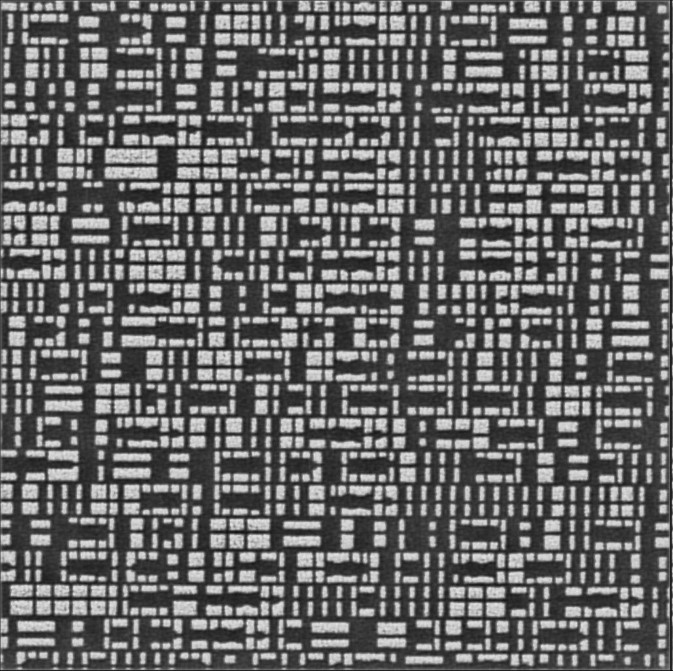}\label{fig:L0_denoised}}%
\hfill
\subfloat[Denoised using $L_2$ loss. PSNR: 24.93 dB]{\includegraphics[width=0.48\textwidth, keepaspectratio]{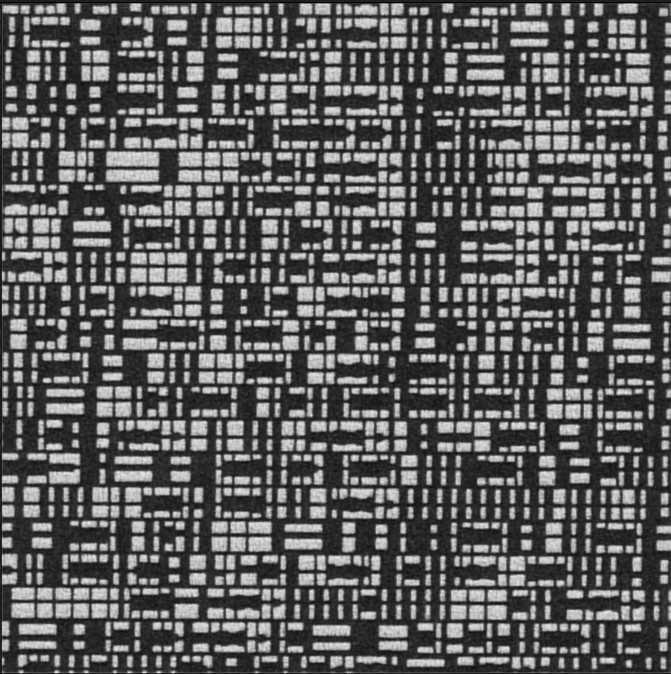}\label{fig:L2_Denoised}}
\caption{Outcome of denoising task.}
\label{fig:denoising_outcome}
\end{figure}

\textbf{\emph{Image Denoising}}. Our SEM imaging setup includes three dwelling times that are 32 $\mu$s/pixel (DT6), 10 $\mu$s/pixel (DT5), and 3.2 $\mu$s/pixel (DT4). We have in total 112 sets of SEM images capturing logical regions of a 28nm node technology. Each set contains three images from a particular position of an IC with three different dwelling time mentioned above.\\
To train the denoising task, we have used all noisy DT4 and DT5 images as input and corresponding DT6 images as ground-truth. We used 85\% images (190) as training and rest (34) as validation.\\
\begin{equation}\label{eq:objective_functions}
\begin{split}
    & L_0 = \frac{1}{N} \sum_{n=1}^{N} (|f(x_i;\theta) - y_i| + \epsilon)^\gamma
    \\
    & L_1 = \frac{1}{N} \sum_{n=1}^{N} |f(x_i;\theta) - y_i|  
    \\
    & L_2 = \frac{1}{N} \sum_{n=1}^{N} (f(x_i;\theta) - y_i)^2 
\end{split}
\end{equation}
where N is the total number of sample, CNN(f) is parameterized by $\theta$, $x_i$ is an input, and $y_i$ is a target. For $L_0$ loss, $\epsilon$ and $\gamma$ have been used to diminish effect of zero gradient and increase focus on foreground area respectively.\\
\hspace*{3mm}Following \cite{DBLP:journals/corr/abs-1803-04189}, we also considered the peak signal-to-noise ratio (PSNR) as the metric for quantitative evaluation. We experimented with three objective functions for the task - $L_0$, $L_1$ and $L_2$ loss (see Equation \ref{eq:objective_functions}). For $L_0$ loss, throughout the training, $\epsilon$ was set to $10^{-8}$ and $\gamma$ was annealed from 2 to 0.\\
\hspace*{3mm}With $L_0$ loss we achieved average PSNR of 22.92 dB whereas with $L_2$ loss we achieved 24.93 dB  (see Fig. \ref{fig:denoising_outcome}). The outcome was not satisfactory with $L_1$ loss.\\
\hspace*{3mm}The number of available training sample (224) is too insufficient for the successful execution for the task. We are hopeful to conduct more experiments in future towards this line of work with more data in hand.

\textbf{\emph{Block Detection}}. Getting accurate outcome from the cell extraction in the block detection unit appears to largely depend on the connected components identified in the images. As mentioned earlier, noise, poor lighting conditions, and excessive rotation all affect the binarization of the images, which results in poor connected component findings. Possible idea for improving the cell extraction procedure is detailed in Section \ref{sec:future-work}.

\begin{figure}[h!]   
\centering
\subfloat[An IC SEM image with improper illumination.]{\includegraphics[width=0.3\textwidth, keepaspectratio]{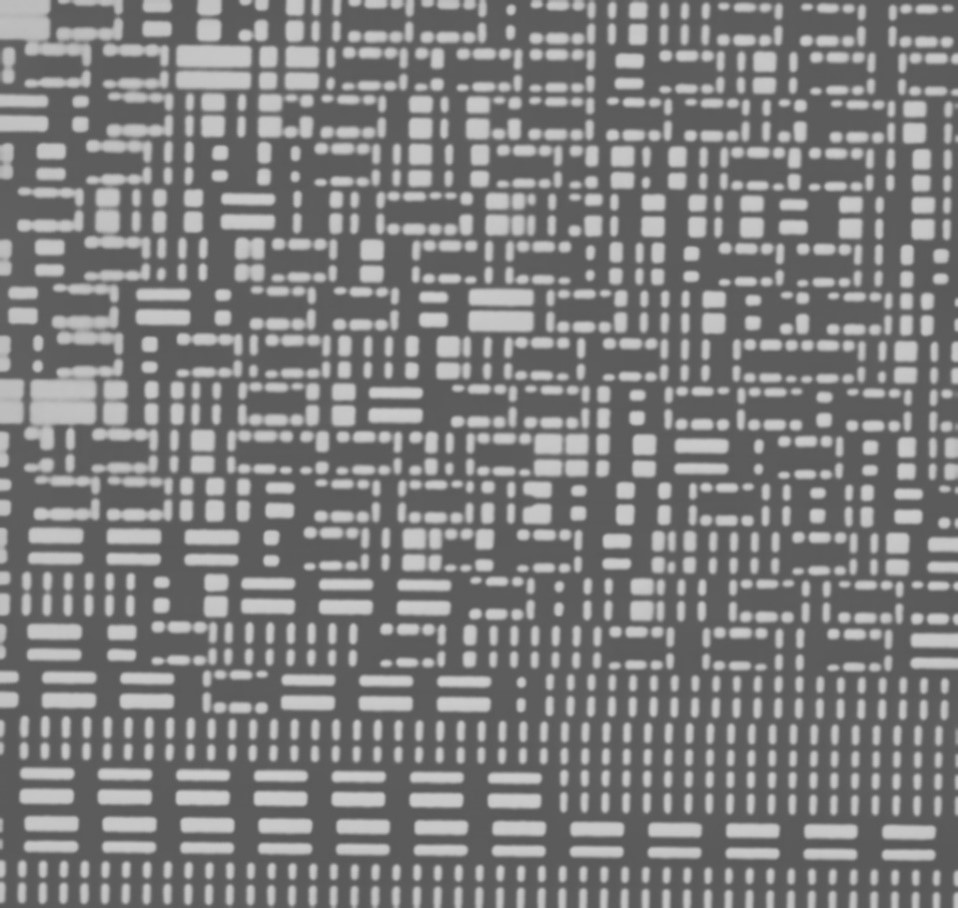}\label{fig:imp-illum}}%
\hfill
\subfloat[Incorrect binarization.]{\includegraphics[width=0.3\textwidth, keepaspectratio]{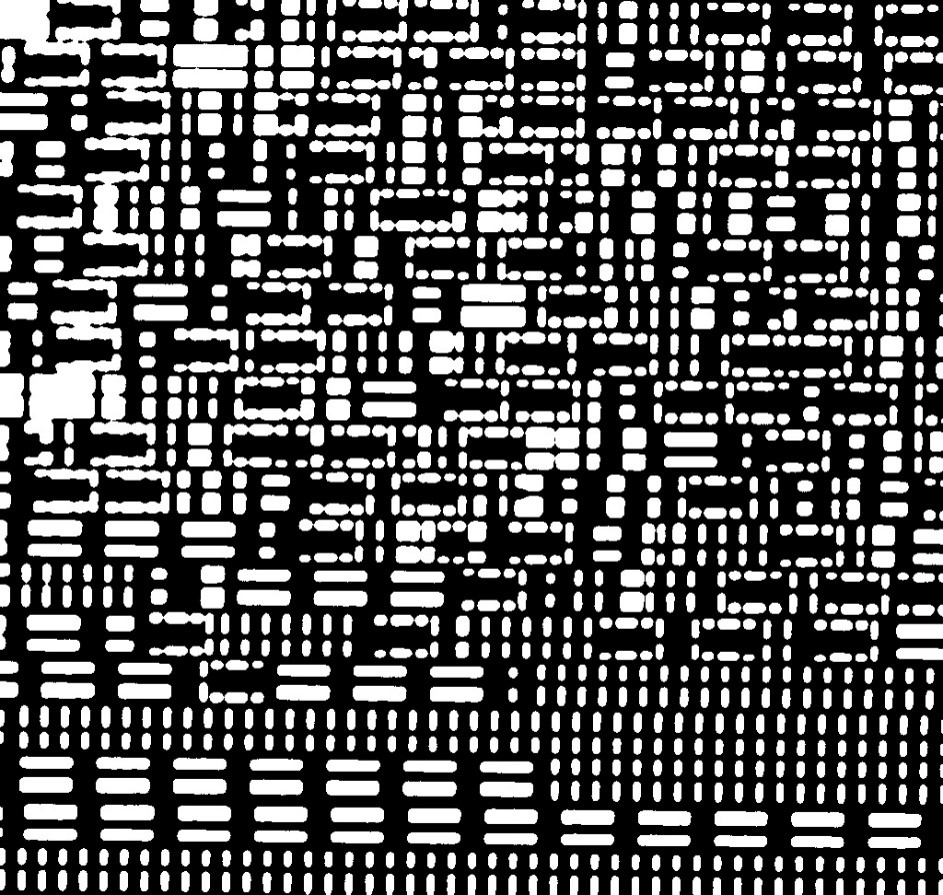}\label{fig:inc-bin}}%
\hfill
\subfloat[Failed identification of connected component.]{\includegraphics[width=0.3\textwidth, keepaspectratio]{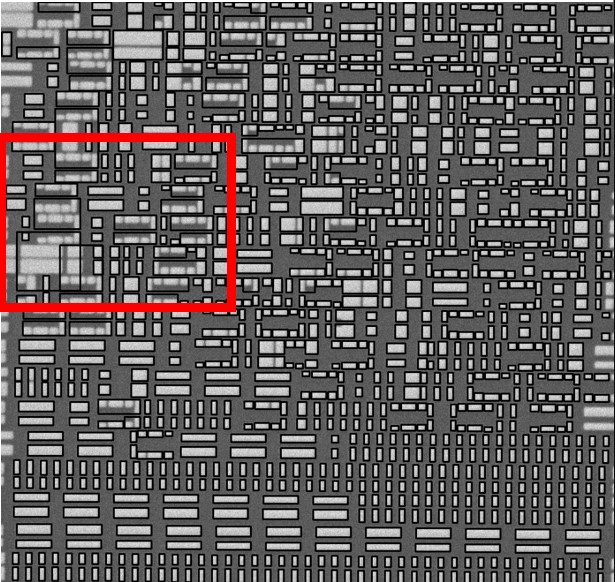}\label{fig:inc-cc}}
\caption{Unsuccessful operations in component findings.}
\label{fig:unsuccessful-processing}
\end{figure}

Our SEM imaging setup includes three dwelling times that are 32 $\mu$s/pixel (DT6), 10 $\mu$s/pixel (DT5), and 3.2 $\mu$s/pixel (DT4). Out of 140 number of IC SEM images used in our experiments, half of them belong to DT5 and the other portion to DT6. The cell extraction process was successful on 116 number of images (82.86\% success rate). The algorithm was not able to operate satisfactory on the images with improper illumination or higher noise level, see Figure \ref{fig:unsuccessful-processing}). The specifications of mentioned images are provided in Table \ref{tab:cell-extraction}. We have plans to overcome these problems in our future work. 

\begin{table}[H]
\centering
\begin{tabular}{ |c|c|c| } 
 \hline
 \textbf{Image Acquisition Setting} & \textbf{Acquired Images} & \textbf{Successful Cell Extraction} \\ 
 \hline
 DT5 & 70 & 58 \\ 
 \hline
 DT6 & 70 & 58 \\ 
 \hline
\end{tabular}
\caption{\label{tab:cell-extraction}The specifications of output images from cell extraction.}
\end{table}

\textbf{\emph{Synthetic Data Generation}}. The extracted cells from the IC SEM images could be used to make an insufficient dataset for the block recognition unit based on its training requirements. Lack of proper diversity among the data samples leads to the inability of the classifier to capture the wide distribution of cell SEM images. To tackle this problem, the synthetic versions of the original cell SEM images are generated using an MSGAN model \cite{synthetic-sem-image}. Due to considering a certain balance and usability among the image samples, seven out of thirteen classes (i.e., types of gates) have been considered in our experiments.
In order to find the difference between the original and the synthetic samples and to provide an analytical evaluation, the standard Jensen Shannon Divergence (JSD) metric \cite{JSD} has been employed. The JSD score for our synthetic data was found to be 0.06 while with random generated noise the JSD score was 0.42. The low JSD value signifies that the generated data was found to be fairly authentic and they were not produced based on a random distribution. The extracted original cell images have been combined with the synthetic images to create a more comprehensive dataset.

\begin{figure}[h!]
\centering
\includegraphics[width=0.7\textwidth, keepaspectratio]{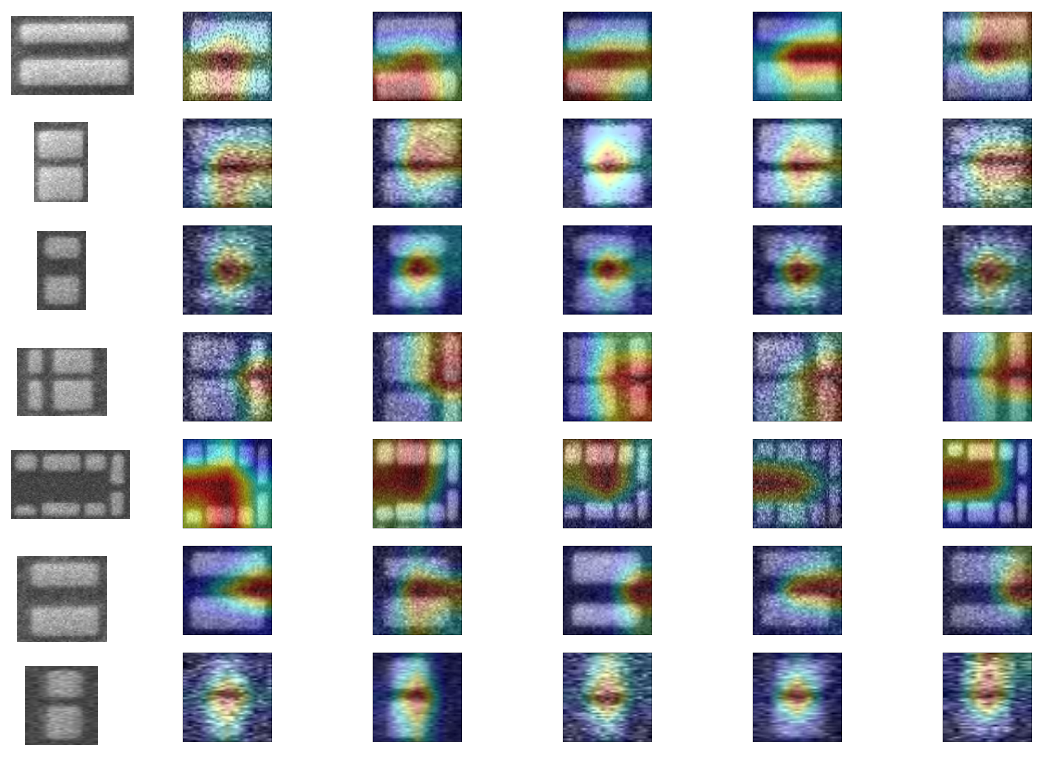}
\caption{The heatmap plot showing the locations activated by the model predictions. The leftmost column contains the actual images of cells.}
\label{fig:gradcam}
\end{figure}

\textbf{\emph{Block Recognition}}. The CNN classifier in the block recognition unit has been trained on seven classes of cell SEM images. The model was able to detect samples with 98\% accuracy score. Our results from this unit along with the block analysis unit are withheld for presentation and they will be shown in a future comprehensive publication on EVHA. In here, our study is focused on the classifier and anomaly detector results on in-class and out-of-class images.\\
To analyze the classifier decisions, a heatmap plot is produced that represents the image locations in which the model predictions have caused the most levels of activation, which is shown in Figure \ref{fig:gradcam}. It can be seen from the figure that region structures as well as the spacing between diffusion regions are important factors in determination of the cell class/type.



\begin{figure}[h!]
    \centering
    \includegraphics[width=0.8\textwidth,keepaspectratio]{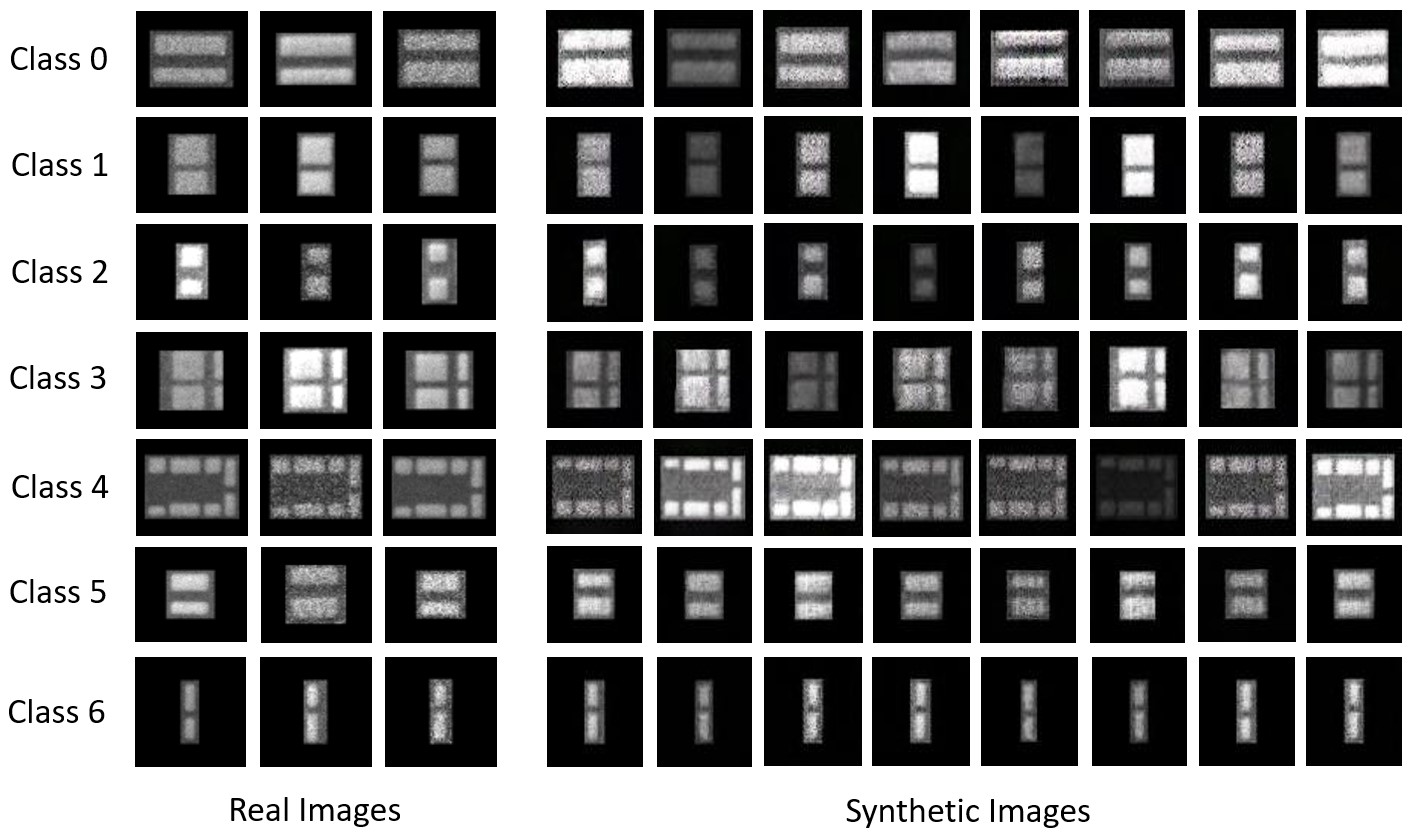}
    \caption{The generated synthetic cell SEM images.}
    \label{fig:synthetic-image}
\end{figure}


\begin{figure}[h!]
\centering
\includegraphics[width=\textwidth, keepaspectratio]{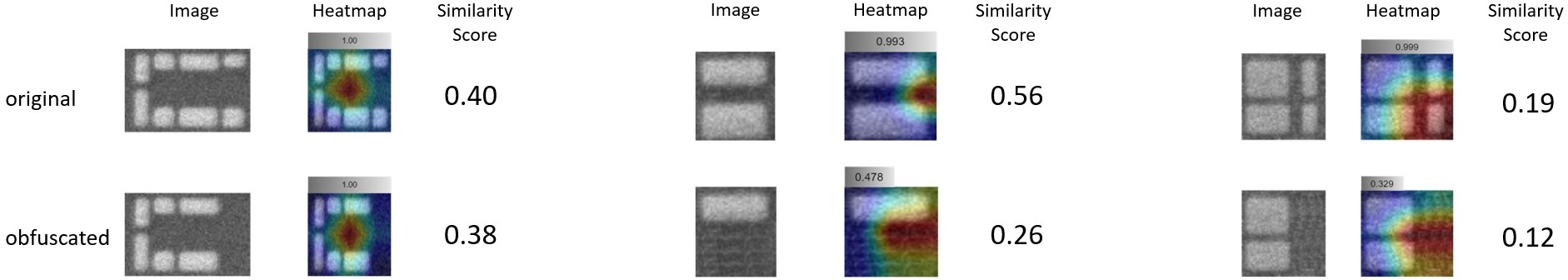}
\caption{The heatmap plot displaying the locations that activate the classifier predictions for a number of out-of-distribution cell SEM image samples. The similarity score has been attained from the anomaly detector. The confidence score is shown on top of each image.}
\label{fig:out-of-distribution}
\end{figure}

In order to examine and observe how the recognition unit operates on anomalous/trojan cells, few modified cell images are fed to the model. Figure \ref{fig:out-of-distribution} shows the predictions and similarity score from classifier and anomaly detector module respectively.  The classifier predicts false positives for anomalous cells with very high confidence. On the other hand, the similarity score from anomaly detector remains slightly higher for clean images than corresponding anomalous ones. The similarity becomes closer when the anomaly is closer to real-word scenario(left pair in figure \ref{fig:out-of-distribution}). The model is not learning enough pattern because we have insufficient and less diverse training data. We are hoping to achieve better result in future with sufficient amount of data.\\
\hspace*{3mm} We still haven't been able to figure out a threshold point to filter out trojan cells. One of the main reason is that we have anomalous sample from multiple clusters/classes of samples. So, intrinsically, we need to solve two problems: image clustering and anomalous sample filtration. This is a comparatively harder problem because threshold levels for anomaly might be different for different clusters of images, and two different kinds of discriminative pattern (class level and anomaly level) need to be learned from same underlying feature.\\
\hspace*{3mm} To the best of our knowledge, this is the first work focused on the self-supervised learning of trojan cells. We believe that this result paves the way for further work in this arena. We have discussed further plan on this research direction in section \ref{sec:future-work}.  


\section{Future Work}\label{sec:future-work}

Our future work will focus on full development and realization of EVHA and presentation of comprehensive and detailed results, analyses, and interpretations. This section introduces few plans for completion and improvement of the system.

\begin{figure}[h!]
    \centering
    \includegraphics[width=\textwidth,keepaspectratio]{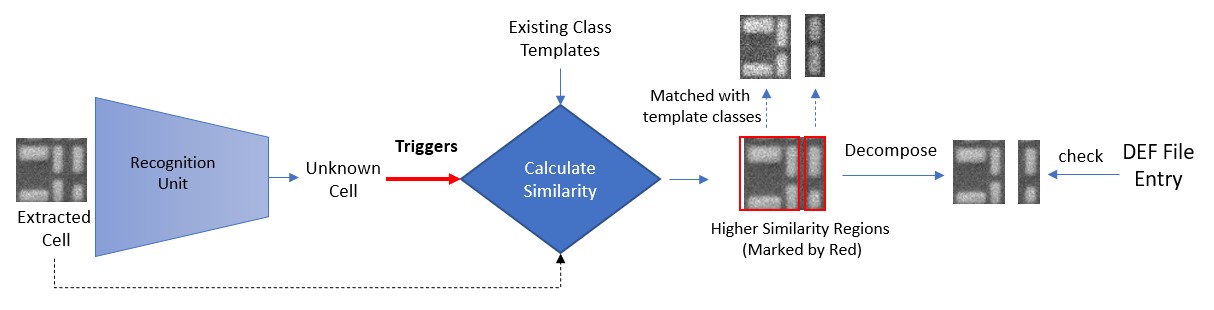}
    \caption{The correction mechanism for cell separator that works based on leveraging prior knowledge.}
    \label{fig:extraction-correction}
\end{figure}

\textbf{Block Detection Unit - Cell Extraction}. One of the main challenges for cell extraction are the separation of single and composite cells based on the distances between consecutive components. Building an intelligent mechanism for analyzing the gaps between components and making differentiation between single and composite cells is a possible solution, but its effectiveness would vary drastically from image to image. So, its system performance can be changed noticeably for different manufacturing technology nodes. As an alternative approach, leveraging prior knowledge as a basis to avoid false negatives can be a more practical and feasible solution, which is in our interest. If a composite cell is extracted incorrectly, then EVHA will classify it as an unknown gate. However, before labelling the composite cell as "malicious", a similarity score will be calculated by comparing each component of the composite cell to the already detected cell classes/types. If a high similarity score is found for each of these components, the composite cell will be decomposed into its singular cells. Next, the DEF file entry will be matched with each decomposed cell. A schematic diagram of this approach is displayed in Figure \ref{fig:extraction-correction}.

Cell library for the corresponding node technology (e.g. 28nm cell library) can be used to label cell types at the primary stage. Later based on the already extracted cells, self-correction can be deployed in the cell extraction system. 

\begin{figure}[h!]
\centering
\includegraphics[width=\textwidth]{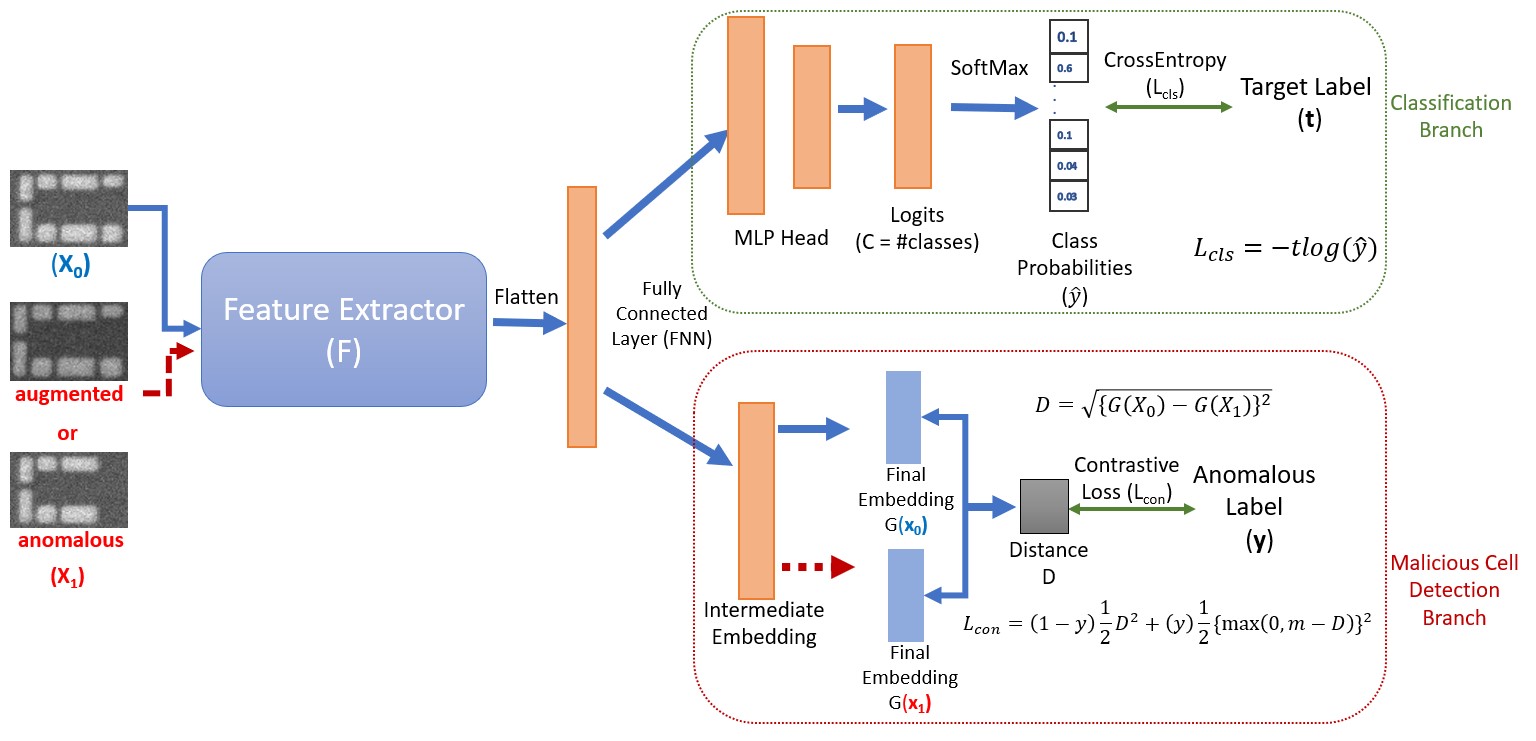}
\caption{Training of the unified cell recognizer. Original view of the image ($X$) and its augmented or anomalous view ($X_1$) are passed through the network. The classification branch predicts the class label from the clean image feature $F(X_0)$. On the other hand, malicious cell detection branch pushes embeddings from image pairs further if $X_1$ is anomalous ($y=1$) otherwise, closer ($y=0$). The network is trained on both classification ($L_{cls}$) and contrastive anomaly loss ($L_{anm}$).}
\label{fig:anomaly-detection-future}
\end{figure}

\textbf{Block Recognition Unit - Anomaly Detection}. Detection of out-of-class/anomaly/trojan cell images (belonging to defect and/or hardware Trojan) is imperative for the system. Data insufficiency is one of the main challenges for our current system as learning both clean and anomalous representation needs enormous amount of data.\\
\hspace*{4mm} The unified framework for clustering and anomalous/trojan cell presence can be a more concrete approach towards the problem. A probable solution towards this problem is depicted in Figure \ref{fig:anomaly-detection-future}. A logical cell image ($x_0$) and its augmented or anomalous view ($x_1$) will be fed to the network consists of a feature extractor F and a classification as well as an anomaly detection branch.\\
\hspace*{4mm} Features extracted by F for both $x_0$ and $x_1$ will be passed through the anomaly detection branch. Embedding from image pair ($G(x_0)$ and $G(x_1)$) will be pushed closer if they are just different views of same image ($y=0$) or further if they are anomalous pair($y=1$ if $x_1$ is anomalous version of $x_0$). The contrastive loss $L_{con}$ serves this purpose (see Figure \ref{fig:anomaly-detection-future}). The classification branch will only act on features from clean image($F(x_0)$). The whole network will be trained jointly on losses from classifier ($L_{cls}$) and anomaly detector ($L_{con}$). We are hoping to show outcome for this multi-task problem in the future.

\section{Conclusion}\label{sec:conclusion}

We have reviewed the framework of the Explainable Vision System for Hardware Testing and Assurance, which is a novel system in the area of physical assurance for the security problem of hardware Trojan detection. It is considered one of the few works in the literature that introduces the concepts and methods from computer vision into the domain of hardware security, specifically for hardware Trojan detection. EVHA intensively leverages advanced object detection and recognition concepts and methods for identifying trojans from cell images and validating cell images upon their collection by a scanning electron microscope. In terms of the number of samples, diversity, quality, and inclusion of both real and synthetic data, the used data stands out from similar works. This review paper studies four major tasks in the framework, including Intelligent Microscopy for Imaging/Delayering (Task 1); Explainable Block/Object Detection and Recognition of IC SEM Images (Task 2); Golden Gate/Circuits Design and Fabrication (Task 3); and Validation and Security Assessment (Task 4). The second task is discussed in more depth through a comprehensive introduction, technical evaluation, and conclusive discussion of its five computing units, namely Image Processing, Block Detection, Block Recognition, Block Analysis, and Decision-Making. It also encompasses our ideas and plans for the full development and realization of EVHA. This work and the proposed EVHA outline will guide the researchers to better understand the SEM imaging domain and the corresponding trojan analysis issues.

\printbibliography

\end{document}